\documentclass[aip,jcp,reprint]{revtex4-2}

\usepackage{amsmath}
\usepackage{graphicx}
\usepackage{dcolumn}
\usepackage{bm}
\usepackage{mathrsfs}
\usepackage{color}
\usepackage{natbib}

\makeatletter
\newcommand{\oast}{\mathbin{\mathpalette\make@circled\ast}}
\newcommand{\make@circled}[2]{%
  \ooalign{$\m@th#1\smallbigcirc{#1}$\cr\hidewidth$\m@th#1#2$\hidewidth\cr}%
}
\newcommand{\smallbigcirc}[1]{%
  \vcenter{\hbox{\scalebox{0.77778}{$\m@th#1\bigcirc$}}}%
}
\makeatother

\begin{document}


\title{Dynamics of Quantum-Classical Systems in Nonequilibrium Environments}

\author{Jeremy Schofield}%
\email{jeremy.schofield@utoronto.ca}
\author{Raymond Kapral}%
\email{r.kapral@utoronto.ca}
\affiliation{Chemical Physics Theory Group, Department of Chemistry, University of Toronto, Toronto, Ontario, Canada, M5S 3H6}

\date{\today}

\begin{abstract}
The dynamics of a quantum system coupled to a classical environment and subject to constraints that drive it out of equilibrium are described. The evolution of the system is governed by the quantum-classical Liouville equation. Rather than evaluating the evolution of the mixed quantum-classical density operator, we derive exact equations of motion for the nonequilibrium average values of a set of operators or variables, along with correlation function expressions for the dissipative coefficients that enter these equations. These equations are obtained by requiring that the exact nonequilibrium averages are equal to local nonequilibrium averages that depend on auxiliary fields whose values satisfy evolution equations obtained using projection operator methods. The results are illustrated by deriving reaction-diffusion equations coupled to fluid hydrodynamic equations for a solution of quantum particles that can exist in two metastable states. Nonequilibrium steady states are discussed along with the reaction rate and diffusion correlation functions that characterize such states.
\end{abstract}

\maketitle

\section{Introduction} \label{sec:intro}

Quantum rate processes in condensed phases are significantly influenced by their environment and the nature of their interactions with it. Reactions such as electron and proton transfer illustrate the importance of coupling with solvent polarization dynamics as a critical aspect of the reaction mechanism\cite{kornyshev97,barbara96,Ali24,warshel1982,marcus1985,Marcus93}. More broadly, environmental effects are crucial in shaping the kinetics of reactions in condensed phases. The microscopic foundation of chemical rate laws and their associated rate coefficients is well understood for systems near equilibrium, with linear response theory and projection operator methods being frequently employed for this purpose ~\cite{reactive-flux,yamamoto60,C78,grote-hynes1980,MST83,0chap-kapral98}. A primary focus of such studies is the calculation of reactive flux correlation functions for quantum rate coefficients. This process requires statistical averaging over quantum equilibrium distributions and the quantum evolution of flux operators, both of which pose challenges in simulations, particularly for large many-body systems.

Some of this difficulty may be alleviated for systems where the environment can be treated classically to a good approximation. It should be noted, however, that quantum-classical approaches are not free from difficulties.~\cite{nielsen01} Depending on how the theory is formulated, one must still statistically sample full quantum or quantum-classical equilibrium distributions, and implement some form of quantum-classical evolution of the coupled quantum and classical subsystems.~\cite{QCreview, mqcl-review18} The environmental degrees of freedom coupled to the quantum subsystem may be treated in various ways. Often the coupling of slowly-varying solvent collective variables to the slowly-varying metastable reactive species variables is most important. In this context, hydrodynamic theories for the dynamics of quantum and quantum-classical systems have been developed.~\cite{burghardt2002,burghardt2011}

Physical systems are often not in equilibrium and are maintained in nonequilibrium steady or dynamical states as a result of the application of external fields or coupling to reservoirs that drive them out of equilibrium~\cite{M11,nitzan24}. Examples include condensed-phase quantum systems where molecular quantum states interact with radiation fields~\cite{Brixner2007}, or quantum molecular junctions coupled to reservoirs that control electron chemical potentials~\cite{GNV15} or temperature gradients across the junction, relevant for electron~\cite{Ratner2007} or heat transport~\cite{Segal2016}. In such cases, one has the additional difficulty of carrying out nonequilibrium quantum or quantum-classical statistical mechanical treatments for both the sampling and dynamics.

In this work, we consider systems whose evolution can be described by quantum-classical dynamics through the quantum-classical Liouville equation~\cite{QCLE2006-review} and are maintained under nonequilibrium conditions by constraints. For this study of reactive quantum-classical systems, we adopt an earlier formulation for nonequilibrium quantum or classical dynamics~\cite{M58, R66, R67, P68, OL79} that has been used in more recent studies of granular materials~\cite{SO93, SO94}, classical Brownian motion~\cite{SO96, SO97} and active particle dynamics~\cite{RSGK20,RSK24}, as well as quantum hydrodynamics~\cite{MG2023}. Rather than solving for the full dynamics of the quantum density operator, this formulation deals with the computation of exact average values of a set of operators using an auxiliary local distribution that depends on constraint fields. The nonlocal and time-dependent constraint fields are obtained by solving coupled dynamical equations.

Section~\ref{sec:QCaverage} gives a general derivation of the time evolution equations for the exact nonequilibrium average values of a set of local operators or phase space variables whose evolution is given by the quantum-classical Liouville equation. These equations express the time derivatives of the average values in terms of the constraint fields or as closed equations for the constraint fields themselves. The formulation also provides nonequilibrium correlation function expressions for transport coefficients that involve statistical averages over the auxiliary local distributions. In Sec.~\ref{sec:RD-hydro} this general formulation is applied to a system containing a solution of s-state quantum molecules that can exist in two metastable states, corresponding to distinct reactive chemical species, in a solvent of classical molecules. The densities of the quantum metastable species are maintained out of equilibrium by reservoirs with given chemical potentials and a fixed temperature. The densities of reactive species are coupled to collective solvent hydrodynamic modes, and a set of coupled reaction-diffusion and hydrodynamic equations is derived, along with the corresponding microscopic expressions for the dissipative transport coefficients. Section~\ref{sec:Rcoord} discusses various possible choices of reaction coordinate and considers the case of a classical solvent coordinate as an illustration of the method. Reactive flux correlation functions are studied in more detail for a dilute solution of quantum molecules using an adiabatic basis, and nonequilibrium steady states are described. The conclusions of the study are given in Sec.~\ref{sec:conc}.

\section{Quantum-classical nonequilibrium average values}\label{sec:QCaverage}
We consider bipartite systems where a quantum subsystem interacts with an environment that is classical in the absence of coupling to the quantum degrees of freedom. The entire system is in contact with reservoirs that maintain it in a nonequilibrium state.  We suppose that the time evolution of the nonequilibrium density operator for this system, $\hat{\rho}(\bm{X},t)$, is given by the quantum-classical Liouville equation,~\cite{alek81,geras81,geras82,kapral99,QCreview}
\begin{align}
\label{eq:qcle}
\partial_t \hat{\rho}(\bm{X},t)&= -\frac{i}{\hbar} [\hat{H}(\bm{X}), \hat{\rho}(\bm{X},t)] \nonumber \\
& \quad - \frac{1}{2}\big(\{\hat{\rho}(\bm{X},t), \hat{H}(\bm{X})\} -\{\hat{H}(\bm{X}), \hat{\rho}(\bm{X},t)\} \big)\nonumber \\
&\equiv -i\hat{\mathcal{L}} \hat{\rho}(\bm{X},t),
\end{align}
where $\bm{X}=(\bm{R},\bm{P})$ are the phase space coordinates of all of the classical degrees of freedom and $i \hat{\mathcal{L}}$ is the quantum-classical Liouville operator.~\cite{ex-fields} The Hamiltonian operator depends on the classical phase space coordinates and is given by the sum of quantum subsystem, coupling operators, and classical environment contributions,
\begin{equation}
\label{eq:H}
\hat{H}(\bm{X}) =\hat{H}_q +  \hat{V}_I(\bm{R}) + H_c(\bm{X}).
\end{equation}

Rather than seeking a solution of the quantum-classical Liouville equation for the density matrix, we focus instead on the computation of the nonequilibrium average values of a set of operators or variables $\widehat{\bm{A}}(\bm{r})$ that may depend on the field point location $\bm{r}$ in the system: $\bm{a}(\bm{r},t)={\rm Tr} [\hat{\rho}(\bm{X},t) \widehat{\bm{A}}(\bm{r})] $, where ${\rm Tr}$ is a trace over the quantum degrees of freedom and the integral is over the classical phase space coordinates. The specific forms of the density operator and ${\rm Tr}$ depend on the physical context and the reservoirs in contact with the system. We denote nonequilibrium average values by lowercase symbols.

The variables of interest depend on the system under study but should vary on time scales that are slow compared to those of microscopic degrees of freedom. The relaxation of a system that is initially displaced from equilibrium may be expressed as a linear combination of the eigenmodes of the Liouville operator. Typically, after an initial short microscopic time, the surviving components of the decomposition of the disturbance consist of a small (but continuous) set of slow modes with small eigenvalues, and the long-time behavior of any dynamical property of the system can then be written as a linear combination of these modes. From these considerations, we see that the set of slow modes $\widehat{\bm{A}}(\bm{r})$ must include the local coarse-grained densities of all conserved variables, such as mass, momentum and energy, but can also include other slow but non-conserved modes.~\cite{other-modes}

\subsection{Local density operators}
The concept of local equilibrium is often used to describe inhomogeneous nonequilibrium systems. In local equilibrium, the state of each small volume of the system is given by thermodynamic relations where variables such as entropy, internal energy, temperature, etc. take values that depend on space and time.~\cite{GM62,GP71} The validity of such a description stems from the observation that the system's relaxation to equilibrium on long-time scales occurs through slowly varying conserved fields, as discussed above.

An exact formulation for the average values of any set of slowly-varying $\widehat{\bm{A}}(\bm{r})$ variables or operators can be given using a local density operator for an open system with $N_\lambda$ classical particles of type $\lambda$ defined by
\begin{equation} 
\label{eq:local-equil}
\hat{\rho}_L(t)=\prod_{\lambda}(N_{\lambda} ! h^{3N_{\lambda}})^{-1}e^{\widehat{\bm{A}}(\bm{r}) \ast \bm{\phi}(\bm{r},t)}/\mathcal{Z}(t),
\end{equation}
where $\mathcal{Z}(t)={\rm Tr}\big[ \prod_{\lambda'}(N_{\lambda'} ! h^{3N_{\lambda'}})^{-1} e^{\widehat{\bm{A}}(\bm{r}) \ast \bm{\phi}(\bm{r},t)}\big]$ and the trace operation is 
\begin{equation}
\label{eq:Trace}
{\rm Tr}[\cdots]=\prod_\lambda \sum_{N_{\lambda}=0}^{\infty{}}{\rm tr} \int d\bm{X}\; \cdots
\end{equation}
where ${\rm tr}$ is a trace over the quantum degrees of freedom and $\int d\bm{X}$ is an integral over the classical phase space. We have chosen to write the local density operator in a generalized form of the grand canonical ensemble since this form will be used in the application we consider. Here and below we use the notation $\hat{B}(\bm{r}) \ast \hat{C}(\bm{r})=\int d\bm{r} \; \hat{B}(\bm{r}) \hat{C}(\bm{r})$ for any local quantities, $\hat{B}(\bm{r})$ and $\hat{C}(\bm{r})$.
The explicit dependence of quantities on the phase space or field point coordinates $\bm{r}$ will sometimes be omitted to simplify the notation. The local density operator is specified by choosing the conjugate $\bm{\phi}$ fields  so that the exact nonequilibrium averages of the $\widehat{\bm{A}}(\bm{r})$ operators under quantum-classical dynamics are equal to their local nonequilibrium averages:~\cite{M58, R66, R67, P68, OL79}
\begin{align}\label{eq:a-def}
\bm{a}(\bm{r},t) &={\rm Tr} \big(\widehat{\bm{A}}(\bm{r}) \hat{\rho}(t) \big)= {\rm Tr} \big(\widehat{\bm{A}}(\bm{r}) \hat{\rho}_L(t)\big)  \\
&\equiv \langle \widehat{\bm{A}}(\bm{r})\rangle_t , \nonumber
\end{align}
where $\langle \cdots \rangle_t$ denotes the average over the local density operator $\hat{\rho}_L(t)$.

The local density operator can be derived using a generalization of the standard Gibbs approach to obtain the equilibrium density~\cite{Gibbs,Reichl2021}. In particular, the form of the local density operator in Eq.~(\ref{eq:local-equil}) can be obtained by maximizing an entropy functional $S(t)$,
\begin{align*}
S(t) = -k_B {\rm Tr} \left[  \rho_L(t) \ln \left( \prod_{\lambda'}(N_{\lambda'} ! h^{3N_{\lambda'}}) \, \rho_L(t) \right) \right],
\end{align*}
with respect to the functional form of $\rho_L(t)$ subject to a set of the locally imposed constraints\cite{OL79} defined in Eq.~(\ref{eq:a-def}).

The $\bm{\phi}(\bm{r},t)$ fields conjugate to the set of dynamical variables $\widehat{\bm{A}}(\bm{r})$ correspond to spatial and time-dependent Lagrange multipliers that enforce the constraint conditions.  As in equilibrium, the Lagrange multipliers and the local nonequilibrium averages of the dynamical variables are  related through functional derivatives of the local equilibrium partition function ${\cal Z}(t)$ and the entropy functional via,
\begin{align*}
\bm{a}(\bm{r},t) &= \frac{\delta \ln {{\cal{Z}} (t)}}{\delta \bm{\phi}  (\bm{r},t)} \\
\bm{\phi}(\bm{r},t) &= -\frac{1}{k_B} \frac{\delta S(t)}{\delta \bm{a}(\bm{r},t)}.
\end{align*}
It can be shown\cite{OL79,GM62,Gaspard_2022} that the entropy production is positive and vanishes in equilibrium where the $\bm{\phi}$ fields are equal to a uniform value.

By way of illustration, consider a single component fluid of classical particles of mass $m$ in contact with reservoirs at the boundaries of the system\cite{LP63,OL79}.  As mentioned above, the minimal set of slow modes of the system ${\bm{A}}(\bm{r})$ must include the local number $N(\bm{r})$, energy $H(\bm{r})$, and momentum $\bm{g}_N(\bm{r})$ densities.  For this set of slow modes, in analogy with equilibrium systems, the conjugate thermodynamic fields are identified to be $\phi_N(\bm{r},t) = \beta (\bm{r},t) \left( \mu (\bm{r},t) - m v^2(\bm{r},t)/2 \right)$, $\phi_E (\bm{r},t) = -\beta (\bm{r},t)$, and $\bm{\phi}_{\bm{g}} (\bm{r},t) = \beta(\bm{r},t) \bm{v}(\bm{r},t)$, where $\beta (\bm{r},t)$ is the local inverse temperature field, $\mu (\bm{r},t)$ is the local chemical potential, and $\bm{v}(\bm{r},t)$ is the local fluid velocity.

At long times, the time-dependence of fluid system is governed by the hydrodynamic equations that describe the evolution of the average densities $\bm{a}(\bm{r},t)$ under the influence of the environment as specified by boundary conditions imposed on the $\bm{a}(\bm{r},t)$ fields or the conjugate thermodynamic fields $\phi (\bm{r},t)$.  For example, thermal gradients can be imposed on the system by specifying nonuniform values of $\beta (\bm{r},t)$ at the boundary, and the response of the system and transport of the local energy, density, and fluid flow can be determined.  Similarly, the flux of particles through a system can be determined as a function of the value of the chemical potential at the boundaries.

Using this formalism, the steady-state values of the local densities and their fluxes can be obtained, and the time evolution of fluctuations from the steady-state values can be determined.  For example, the light-scattering spectrum and long-ranged static correlations in systems maintained in a steady state with linear temperature or velocity profiles have been studied\cite{Machta82}.
Section~\ref{sec:Rcoord} provides another example of the computation of steady-state densities for a reaction-diffusion system.

Although more general forms of the local equilibrium density operator can be formulated to include constraints placed on the nonequilibrium averages of multilinear densities as well\cite{SO94}, we will not consider these complications here since they only account for smaller mode-coupling corrections to the transport coefficients that appear in the dynamics of averages of linear densities.

Next, we derive the equations of motion for the average $\bm{a}(\bm{r},t)$ fields and their conjugate $\bm{\phi}(\bm{r},t)$ fields whose solutions will yield the desired values of these quantities. While the constraint condition in Eq.~(\ref{eq:a-def}) imposes equality between the full and local equilibrium averages of the chosen $\widehat{\bm{A}}(\bm{r})$ variables, the equality does not hold for the averages of other fields. Consequently, to derive the equations of motion for the $\bm{a}(\bm{r},t)$ and $\bm{\phi}(\bm{r},t)$ fields, we need to express the full quantum-classical density operator in terms of the local density operator. This expression is derived in below in Sec.~\ref{subsec:density-operators}, while equations of motion are given in Sec.~\ref{subsec:evola}. 

\subsection{Relation between exact and local density operators}\label{subsec:density-operators}

To obtain the relation between exact and local density operators, we introduce a projection operator $\hat{\mathcal{P}}^\dagger(t)$ and its complement $\hat{\mathcal{Q}}^\dagger(t) = 1 - \hat{\mathcal{P}}^\dagger(t)$ that projects the density operator onto  the local equilibrium density operator and satisfies
\begin{align}
\label{eq:density-split}
\hat{\rho} (t) = \hat{\mathcal{P}}^\dagger(t) \hat{\rho}(t) + \hat{\mathcal{Q}}^\dagger(t) \hat{\rho}(t)
= \hat{\rho}_L(t) + \hat{\mathcal{Q}}^\dagger(t) \hat{\rho}(t).
\end{align}
A projection operator ${\mathcal P}^\dagger (t)$ with these properties can be constructed using the following considerations: We note that the local equilibrium density operator depends explicitly on time since it is a functional of the time-dependent thermodynamic fields. It follows that the time evolution of this density is governed by
\begin{align}
\label{eq:time-deriv-rhoL}
\partial_t \hat{\rho}_L(t) &= \frac{\delta \hat{\rho}_L(t)}{\delta \bm{\phi}(\bm{r},t)} \ast \partial_t {\bm{\phi}}(\bm{r},t).
\end{align}
To evaluate the functional derivative of $\hat{\rho}_L(t)$,
we make use of the operator identity,
\begin{equation*}
e^{x(\hat{B}+\hat{C})} = e^{x\hat{B}} + \int_0^1 dx' \;e^{x'(\hat{B}+\hat{C})} \hat{C} e^{(x-x') \hat{B}}. \end{equation*}
Letting $\hat{B}=\widehat{\bm{A}} \ast \bm{\phi}$ and $\hat{C}=\widehat{\bm{A}} \ast \delta \bm{\phi}$
we have to linear order,
\begin{equation*}
e^{\widehat{\bm{A}}\ast (\bm{\phi} +\delta \bm{\phi})} \approx  e^{\widehat{\bm{A}}\ast \bm{\phi}} + \int_0^1 dx' \;e^{x'\widehat{\bm{A}}\ast \bm{\phi}} \widehat{\bm{A}} e^{-x' \widehat{\bm{A}}\ast \bm{\phi}} \ast \delta \bm{\phi} \, e^{\widehat{\bm{A}}\ast \bm{\phi}},\nonumber \\
\end{equation*}
that yields
\begin{equation}
\label{eq:functional-derivative}
\frac{\delta \hat{\rho}_L(t)}{\delta \bm{\phi}(\bm{r},t)} = \overline{\bm{A}} (\bm{r}) \hat{\rho}_L(t)= \hat{\rho}_L(t)\overline{\bm{A}}^\dagger (\bm{r}) ,
\end{equation}
which involves $\overline{\bm{A}}(\bm{r})$ and its Hermitian conjugate $\overline{\bm{A}}^\dagger(\bm{r})$.

The overline is used to denote the nonequilibrium analog of a Kubo-transformed variable~\cite{kubo1957} that, for any Hermitian operator $\hat{O}(\bm{r}) = \hat{O}^\dagger (\bm{r})$, is given by
\begin{equation}
\label{eq:noneqKubo}
\overline{O}(\bm{r})=\int_0^1 dx \; e^{x \widetilde{\bm{A}} \ast \bm{\phi}} \widetilde{O}(\bm{r}) e^{-x \widehat{\bm{A}} \ast \bm{\phi}} ,
\end{equation}
along with a corresponding expression for $\overline{O}^\dagger(\bm{r})$. Here and below we use the notation $\widetilde{O}(\bm{r})=\hat{O}(\bm{r}) -\langle \hat{O}(\bm{r}) \rangle_t$.

Similarly, from the constraint condition in Eq.~(\ref{eq:a-def}) we see that the average value $\bm{a}(\bm{r},t)$ is also a functional of the $\bm{\phi}$ fields with functional derivative,
\begin{equation*}
\frac{\delta \bm{a}(\bm{r}_1,t)}{\delta \bm{\phi}(\bm{r}_2,t)}=\langle \widetilde{\bm{A}}(\bm{r}_1) \overline{\bm{A}}(\bm{r}_2)\rangle_t =\langle \overline{\bm{A}}^\dagger(\bm{r}_2) \widetilde{\bm{A}}(\bm{r}_1)\rangle_t .
\end{equation*}
The equivalent forms on the right of this equation are a consequence of properties of the correlation functions of Hermitian operators: $\langle \widetilde{O}_\alpha (\bm{r}_1) \overline{O}_{\beta}(\bm{r}_2) \rangle_t = \langle \overline{O}^\dagger_\alpha (\bm{r}_1) \widetilde{O}_\beta(\bm{r}_2) \rangle_t =  \langle \widetilde{O}_\beta (\bm{r}_2) \overline{O}_\alpha (\bm{r}_1) \rangle_t$.
The inverse of this relation is
\begin{equation*}
\frac{\delta \bm{\phi}(\bm{r}_2,t)}{\delta \bm{a}(\bm{r}_3,t)}=\langle \widetilde{\bm{A}} \overline{\bm{A}} \rangle^{-1}_t(\bm{r}_2,\bm{r}_3)=\langle  \overline{\bm{A}}^\dagger \widetilde{\bm{A}}\rangle^{-1}_t(\bm{r}_3,\bm{r}_2) .
\end{equation*}

Inserting Eq.~(\ref{eq:functional-derivative}) into Eq.~(\ref{eq:time-deriv-rhoL}) yields the results
\begin{align*}
\partial_t \hat{\rho}_L(t) &= \left( \overline{\bm{A}}(\bm{r})  \ast \partial_t \bm{\phi}(\bm{r},t) \right) \hat{\rho}_L(t) \nonumber \\
&= \hat{\rho}_L(t) \left( \overline{\bm{A}}^\dagger (\bm{r})  \ast \partial_t \bm{\phi}(\bm{r},t) \right)
\end{align*}
Note that the local equilibrium density is expected to evolve slowly since the thermodynamic fields depend on time through their functional dependence on the slow densities $\bm{a}(\bm{r},t)$.

These considerations suggest defining the projector operator $\hat{\mathcal{P}}^\dagger(t)$
that acts on an arbitrary density operator $\hat{f}(t)$ and yields an operator proportional to the local density. Hence, we define the projection operator\cite{R66,OL79},
\begin{eqnarray}
\label{eq:Pdag-proj}
\hat{\mathcal{P}}^\dagger(t) \hat{f}(t)&=& {\rm Tr}\big(\hat{f}(t)\big) \hat{\rho}_L(t)  \\
&& +{\rm Tr}\big(\hat{f}(t)\widetilde{\bm{A}}(\bm{r}_1)\big) \ast \langle \widetilde{\bm{A}} \overline{\bm{A}} \rangle^{-1}_t(\bm{r}_1,\bm{r}_2)\nonumber \\
&&\qquad \ast \frac{1}{2}\Big(\overline{\bm{A}}(\bm{r}_2)\hat{\rho}_L(t) + \hat{\rho}_L(t) \overline{\bm{A}}^\dagger(\bm{r}_2)\Big),\nonumber
\end{eqnarray}
written in a symmetric form that emphasizes the Hermitian properties of the projector. The form of the projector in Eq.~(\ref{eq:Pdag-proj}) can also be obtained from the alternate expression,
\begin{equation*}
\hat{\mathcal{P}}^\dagger(t) \hat{f}(t)= {\rm Tr}\big(\hat{f}(t)\big) \hat{\rho}_L(t) +\frac{\delta \hat{\rho}_L(t)}{\delta \bm{a}(\bm{r},t)}\ast {\rm Tr}\big(\widehat{\bm{A}}(\bm{r}) \hat{f}(t)\big).
\end{equation*}

If this projector acts on the exact density $\hat{\rho}(t)$, we have $\hat{\mathcal{P}}^\dagger(t) \hat{\rho}(t)=\hat{\rho}_L(t)$, since ${\rm Tr}\big(\hat{\rho}(t)\widetilde{\bm{A}}(\bm{r}_1)\big)=0$ by construction. Furthermore, by including additional terms in the projection operator involving the variational derivative of $\hat{\rho}_L (t)$ with respect to the densities, we obtain the result that ${\mathcal P}^\dagger(t) \partial_t \hat{\rho}_L(t) = \partial_t \hat{\rho}_L(t)$, as can be verified by applying the projection operator definition in Eq.~(\ref{eq:Pdag-proj}) to the time derivative of $\hat{\rho}_L(t)$ given in Eq.~(\ref{eq:time-deriv-rhoL}).  Since the constraint condition implies that ${\rm{Tr}} \big[ \partial_t \hat{\rho}(t) \bm{\widehat{A}}(\bm{r}) \big] = {\rm{Tr}} \big[ \partial_t \hat{\rho}_L(t) \bm{\widehat{A}}(\bm{r}) \big]$, we obtain ${\mathcal P}^\dagger (t) \partial_t \hat{\rho}(t) = \partial_t \hat{\rho}_L (t) = \partial_t \, {\mathcal P}^\dagger (t) \hat{\rho}(t)$.

To complete the decomposition of $\hat{\rho}(t)$ in terms of $\hat{\rho}_L(t)$, we require the solution of the evolution of ${\mathcal Q}^\dagger (t) \hat{\rho}(t)$ whose equation of motion is
\begin{align*}
\partial_t \hat{\mathcal{Q}}^\dagger(t) \hat{\rho}(t) &=
\partial_t \hat{\rho}(t) - {\mathcal P}^\dagger (t) \partial_t \hat{\rho}(t) = {\mathcal Q}^\dagger (t) \partial_t \hat{\rho}(t) \\
&=
-\hat{\mathcal{Q}}^\dagger(t) i\hat{\mathcal{L}}\hat{\rho}_L(t) -\hat{\mathcal{Q}}^\dagger(t) i \hat{\mathcal{L}} \hat{\mathcal{Q}}^\dagger(t) \hat{\rho}(t).
\end{align*}
The formal solution of this equation is
\begin{align*}
\hat{\mathcal{Q}}^\dagger(t) \hat{\rho}(t) = &\hat{U}^\dagger_Q(t,0)\hat{\mathcal{Q}}^\dagger(0) \hat{\rho}(0) 
\\
&-\int_0^t dt_1 \; \hat{U}^\dagger_Q(t,t_1) \hat{\mathcal{Q}}^\dagger(t_1) i\hat{\mathcal{L}}  \hat{\rho}_L(t_1), \nonumber
\end{align*} 
where $\hat{U}^\dagger_Q(t,t_1)$ is the time-ordered exponential
\begin{equation*}
\hat{U}^\dagger_Q(t,t_1)=\mathcal{T}_+ \exp\Big(-\int_{t_1}^t dt' \; \hat{\mathcal{Q}}^\dagger(t')i\hat{\mathcal{L}} \Big),
\end{equation*}
with smaller time arguments appearing to the right of larger time arguments. 
The desired relation between the two density operators is, therefore,
\begin{eqnarray}\label{eq:rho-rhoL}
\hat{\rho}(t)&=& \hat{\rho}_L(t) +\hat{U}^\dagger_Q(t,0)\hat{\mathcal{Q}}^\dagger(0) \hat{\rho}(0) \\
&&-\int_0^t dt_1 \; \hat{U}^\dagger_Q(t,t_1) \hat{\mathcal{Q}}^\dagger(t_1) i\hat{\mathcal{L}}  \hat{\rho}_L(t_1) . \nonumber
\end{eqnarray}

In the following, it will be useful to consider the Hermitian conjugate $\hat{\mathcal{P}}(t)$ of the projector $\hat{\mathcal{P}}^\dagger(t)$ defined through the relation,
\begin{eqnarray*}
{\rm Tr} \big[ \hat{O}_\alpha \big( \hat{\mathcal{P}}^\dagger(t) \hat{f}(t) \big) \big] &=
{\rm Tr} \big[  \hat{f}(t) \big( \hat{\mathcal{P}}(t) \hat{O}_\alpha \big)  \big]
\end{eqnarray*}
and given by
\begin{equation*}
\hat{\mathcal{P}}(t) \widetilde{O}_\alpha =  \langle \widetilde{O}_\alpha \overline{\bm{A}}(\bm{r}_1) \rangle_t \ast
 \langle \widetilde{\bm{A}} \overline{\bm{A}} \rangle^{-1}_t(\bm{r}_1,\bm{r}_2) \ast \widetilde{\bm{A}}(\bm{r}_2) .
\end{equation*}
It follows that the complement of this projector, $\hat{\cal{Q}}(t) = 1 - \hat{\cal{P}}(t)$, removes the correlation of an operator $\hat{O}$ with the set of slow densities $\widehat{\bm{A}}(\bm{r})$ since $\langle ( \hat{\cal{Q}}(t) \hat{O} ) \overline{\bm{A}}(\bm{r}) \rangle_t = 0$.

\subsection{Evolution equations for $\bm{a}(\bm{r},t)$ and $\bm{\phi}(\bm{r},t)$}\label{subsec:evola}
The evolution equation for $\bm{a}(\bm{r},t)$ follows from the time derivative of the definition $\bm{a}(\bm{r},t)$ in Eq.~(\ref{eq:a-def}):
\begin{equation}
\label{eq:dadt}
\partial_t \bm{a}(\bm{r},t)= {\rm Tr} \big(\hat{\rho}(t) \hat{\bm{\mathcal{J}}}_{A}(\bm{r})\big),
\end{equation}
with $\hat{\bm{\mathcal{{J}}}}_{A}(\bm{r}) =  i\hat{\mathcal{L}} \widehat{\bm{A}}(\bm{r})$ giving
\begin{equation*}
\hat{\bm{\mathcal{J}}}_{A}(\bm{r})=\frac{i}{\hbar} [\hat{H}, \widehat{\bm{A}}(\bm{r})] + \frac{1}{2}\big( \{\widehat{\bm{A}}(\bm{r}), \hat{H}\} - \{\hat{H},\widehat{\bm{A}}(\bm{r})\} \big).
\end{equation*}
Then, substituting Eq.~(\ref{eq:rho-rhoL}) into Eq.~(\ref{eq:dadt}) we obtain the evolution equation,
\begin{eqnarray}
\label{eq:a-evol1}
&&\partial_t \bm{a}(\bm{r},t)= \langle \hat{\bm{\mathcal{J}}}_{A}(\bm{r}) \rangle_t + {\rm Tr} \big( \hat{\bm{\mathcal{J}}}_{A}(\bm{r}) \hat{U}^\dagger_Q(t,0)\hat{\mathcal{Q}}^\dagger(0) \hat{\rho}(0) \big) \nonumber \\
&&\quad -\int_0^t dt_1 \; {\rm Tr} \big( \hat{\bm{\mathcal{J}}}_{A}(\bm{r})\hat{U}^\dagger_Q(t,t_1) \hat{\mathcal{Q}}^\dagger(t_1) i\hat{\mathcal{L}} \hat{\rho}_L(t_1)\big) .
\end{eqnarray}

In Eq.~(\ref{eq:a-evol1}) we can move the evolution operator onto the first flux by making use of the relation
\begin{equation*}
{\rm Tr} \Big(\hat{B} \, \hat{U}^\dagger_Q(t,t_1) \hat{C}(t_1) \Big)={\rm Tr} \Big( (\hat{U}_Q(t,t_1)\hat{B}) \, \hat{C}(t_1) \Big)
\end{equation*}
where
\begin{equation*}
\hat{U}_Q(t,t_1)=\mathcal{T}_- \exp \Big(\int_{t_1}^t dt' \;i\hat{\mathcal{L}} \hat{\mathcal{Q}}(t') \Big),
\end{equation*}
and the time ordering operator $\mathcal{T}_-$ signifies that larger time arguments appear to the right of smaller time arguments. This allows us to write
\begin{eqnarray*}
&&{\rm Tr} \big( \hat{\bm{\mathcal{J}}}_{A}(\bm{r})\hat{U}^\dagger_Q(t,t_1) \hat{\mathcal{Q}}^\dagger(t_1) i\hat{\mathcal{L}} \hat{\rho}_L(t_1)\big)=\\
&& \qquad \qquad {\rm Tr} \big( \hat{\bm{\mathcal J}}_{A,t}(\bm{r},t_1,t) \hat{\mathcal{Q}}^\dagger(t_1) i\hat{\mathcal{L}} \hat{\rho}_L(t_1)\big) \nonumber
\end{eqnarray*}
where $\hat{\bm{{\mathcal J}}}_{A,t}(\bm{r},t_1,t) =  U_{Q}(t_1,t) \hat{\bm{\mathcal{J}}}_{A,t}(\bm{r})$ with the projected flux operator defined by $\hat{\bm{\mathcal{J}}}_{A,t}(\bm{r}) \equiv \hat{\mathcal{Q}}(t)\hat{\bm{\mathcal{J}}}_{A}(\bm{r})$ indicated by the subscript $t$.

To express the memory term as an average over the local density operator, we need an expression for $ i\hat{\mathcal{L}} \hat{\rho}_L(t)$. The quantum-classical Liouville operator acting on the local density operator consists of two parts;  the first consists of the commutator of the local density operator with the Hamiltonian operator, and the second involves the anti-symmetrized  Poisson bracket of the local density with the Hamiltonian operator,
\begin{eqnarray*}
i\hat{\mathcal{L}} \hat{\rho}_L(t)&=&\frac{1}{\mathcal{Z}} \prod_{\lambda}(N_{\lambda} ! h^{3N_{\lambda}})^{-1}
\Big( \frac{i}{\hbar} [\hat{H}, e^{\widehat{\bm{A}} \ast \bm{\phi}}] \\
&+&\frac{1}{2} \big(  \{e^{\widehat{\bm{A}} \ast \bm{\phi}}, \hat{H}\} -  \{\hat{H},e^{\widehat{\bm{A}} \ast \bm{\phi}}\} \big) \Big) \nonumber .
\end{eqnarray*}
The commutator in this expression can be evaluated to give
\begin{eqnarray*}
[\hat{H}, e^{\widehat{\bm{A}} \ast \bm{\phi}}]&=& -\int_0^1 dx \; \frac{d}{dx}\big(  e^{x\widehat{\bm{A}} \ast \bm{\phi}}\hat{H}  e^{(1-x)\widehat{\bm{A}} \ast \bm{\phi}}\big)  \nonumber \\
&=& \int_0^1 dx \; \big( e^{x\widehat{\bm{A}} \ast \bm{\phi}} [\hat{H},\widehat{\bm{A}}] e^{-x\widehat{\bm{A}} \ast \bm{\phi}}  \big)  \ast \bm{\phi} \; e^{\widehat{\bm{A}} \ast \bm{\phi}}\nonumber\\
&\equiv& \overline{[\hat{H},\widehat{\bm{A}}]} \ast \bm{\phi}\; e^{\widehat{\bm{A}} \ast \bm{\phi}}.
\end{eqnarray*}
The terms involving the Poisson brackets can be rewritten using the identity
\begin{eqnarray*}
\partial_{\bm{X}} e^{\widehat{\bm{A}} \ast \bm{\phi}}&=& \int_0^1 dx \; \big( e^{x\widehat{\bm{A}} \ast \bm{\phi}} (
\partial_{\bm{X}} \widehat{\bm{A}}) e^{-x\widehat{\bm{A}} \ast \bm{\phi}}  \big)   \ast \bm{\phi} \; e^{\widehat{\bm{A}} \ast \bm{\phi}}\nonumber\\
&=&\overline{( \partial_{\bm{X}}
\widehat{\bm{A}}) } \ast \bm{\phi}\; e^{\widehat{\bm{A}} \ast \bm{\phi}},
\end{eqnarray*}
and hence the phase space derivative of the local density is
\begin{equation*}
\partial_{\bm{X}}  \hat{\rho}_L(t) = \left( \overline{( \partial_{\bm{X}}\widehat{\bm{A}}) } \ast \bm{\phi}\right) \hat{\rho}_L(t)  = \hat{\rho}_L(t)  \left( \overline{( \partial_{\bm{X}}\widehat{\bm{A}}) }^\dagger \ast \bm{\phi} \right).
\end{equation*}
Combining these results, we obtain
\begin{equation*}
i\hat{\mathcal{L}} \hat{\rho}_L(t)= \frac{1}{2}\Big(\overline{\bm{J}}_{A} (\bm{r})  \hat{\rho}_L(t) +\hat{\rho}_L(t) \overline{\bm{J}}^\dagger_{A} (\bm{r})\Big) \ast \bm{\phi} (\bm{r},t),
\end{equation*}
where the transformed flux $\overline{\bm{J}}_{A} (\bm{r})$ is
\begin{equation*}
\overline{\bm{J}}_{A} (\bm{r}) =  \frac{i}{\hbar} \overline{ [\hat{H},\widehat{\bm{A}}(\bm{r})]}
- \partial_{\bm{R}} \hat{H} \cdot \overline{\partial_{\bm{P}} \widehat{\bm{A}}(\bm{r}) }  +
\frac{\bm{P}}{M} \cdot \overline{\partial_{\bm{R}} \widehat{\bm{A}}(\bm{r})},
\end{equation*}
and $\overline{\bm{J}}^\dagger_{A} (\bm{r})$ is defined analogously.

We need one further simplification of the projected time derivative of the local density, $\hat{\mathcal{Q}}^\dagger (t) i{\cal L} \hat{\rho}_L(t)$. Using the definitions of the projection operators, we have
\begin{eqnarray*}
\hat{\mathcal{Q}}^\dagger(t) \overline{\bm{J}}_A(\bm{r}) \hat{\rho}_L(t)  &=&
\big( \overline{\mathcal{Q}}(t)\overline{\bm{J}}_{A}\big) \hat{\rho}_L(t) \equiv  \overline{\bm{\mathcal{J}}}_{A,t} \hat{\rho}_L(t), \\
  \hat{\mathcal{Q}}^\dagger(t)\Big( \hat{\rho}_L(t) \overline{\bm{J}}^\dagger_{A}\Big)&=&  \hat{\rho}_L(t) \big(\overline{Q}^\dagger(t)\overline{\bm{J}}^\dagger_{A} \big) \equiv \hat{\rho}_L(t)  \overline{\bm{\mathcal{J}}}^\dagger_{A,t} ,
\end{eqnarray*}
where we have introduced the new projectors $\overline{Q}(t) = 1 - \overline{P}(t)$ and $\overline{Q}^\dagger (t) = 1 - \overline{P}^\dagger (t)$ with
\begin{align*}
\overline{P}(t)\overline{\bm{J}}_{A} & =
\langle \overline{\bm{J}}_{A} \rangle_t + \overline{\bm{A}} \ast \langle  \widetilde{\bm{A}} \overline{\bm{A}} \rangle^{-1}_t \ast \langle \widetilde{\bm{A}} \overline{\bm{J}}_{A}  \rangle_t
\\
\overline{P}^\dagger (t)\overline{\bm{J}}^\dagger_{A} &=
\langle \overline{\bm{J}}^\dagger_{A} \rangle_t +  \langle  \overline{\bm{J}}^\dagger_{A} \widetilde{\bm{A}}  \rangle_t \ast \langle  \overline{\bm{A}}^\dagger \widetilde{\bm{A}} \rangle^{-1}_t  \ast \overline{\bm{A}}^\dagger
.
\end{align*}

The evolution equation~(\ref{eq:a-evol1}) can now be written as
\begin{align}
\label{eq:Qa-phi}
\partial_t \bm{a}(\bm{r},t)&= \langle  \hat{\bm{\mathcal{J}}}_{A}(\bm{r}) \rangle_t + \bm{I}(\bm{r},t)  \\
& \qquad -\int_0^t dt_1 \; \bm{\Gamma}(\bm{r}, \bm{r}_1, t, t_1) \ast \bm{\phi}(\bm{r}_1,t_1) ,\nonumber
\end{align}
where the dissipative matrix $\bm{\Gamma}$ is defined to be
\begin{eqnarray*}
\bm{\Gamma}(\bm{r}, \bm{r}_1, t , t_1) &=& \big\langle  \hat{\bm{{\mathcal J}}}_{A,t}(\bm{r},t_1,t)\overline{\bm{\mathcal{J}}}_{A,t_1}(\bm{r}_1)   \big\rangle^{\rm sym}_{t_1} \\
&\equiv&
 \frac{1}{2} \Big( \big\langle  \hat{\bm{{\mathcal J}}}_{A,t}(\bm{r},t_1,t)\overline{\bm{\mathcal{J}}}_{A,t_1}(\bm{r}_1)   \big\rangle_{t_1} \nonumber
  \\
 && \quad + \big\langle \overline{\bm{\mathcal{J}}}^\dagger _{A,t_1}(\bm{r}_1) \hat{\bm{{\mathcal J}}}_{A,t}(\bm{r},t_1,t)   \big\rangle_{t_1} \Big) . \nonumber
\end{eqnarray*}
In Eq.~(\ref{eq:Qa-phi}), the term depending on the initial density operator is $\bm{I}(\bm{r},t) ={\rm Tr} \big(  \big( \hat{U}_Q(t,0) \hat{\bm{\mathcal{J}}}_{A}(\bm{r}) \big) \, \hat{\mathcal{Q}}^\dagger(0) \hat{\rho}(0) \big)$, and the random force operators evolve with the projected evolution operator $\hat{U}_Q$.

An equation of motion for $\bm{\phi}(\bm{r},t)$ can be obtained by first making use of Eq.~(\ref{eq:time-deriv-rhoL}) to get
\begin{equation}\label{eq:adot-phidot}
\partial_t \bm{a}(\bm{r}_1,t)=\langle \widetilde{\bm{A}}(\bm{r}_1) \overline{\bm{A}}(\bm{r})\rangle_t \ast \partial_t{\bm{\phi}}(\bm{r},t),
\end{equation}
and then using it in Eq.~(\ref{eq:Qa-phi}). Letting $\bm{K}_t(\bm{r}_1,\bm{r})=\langle \widetilde{\bm{A}}(\bm{r}_1) \overline{\bm{A}}(\bm{r})\rangle_t$, we find
\begin{align}\label{eq:Qphi}
\partial_t \bm{\phi}(t) &= \bm{K}_t^{-1} \ast \big(\langle \hat{\bm{\mathcal{J}}}_A \rangle_t +\bm{I}(t)\big) \\
 &\quad -\int_0^t dt_1 \; \bm{K}_t^{-1} \ast \bm{\Gamma}(t, t_1) \ast \bm{\phi}(t_1), \nonumber
\end{align}
where the spatial dependence has been suppressed for simplicity.
The set of integro-differential equations~(\ref{eq:Qphi}), although complicated, is exact and can be solved self-consistently to give the values of the $\bm{\phi}$ fields needed to compute the exact nonequilibrium average values of $\widehat{\bm{A}}(\bm{r})$ given in Eq.~(\ref{eq:a-def}).

Equations~(\ref{eq:Qa-phi}) and (\ref{eq:Qphi}) are the exact general evolution equations that will form the basis of the subsequent calculations in this paper.  While the set of densities of operators $\widehat{\bm{A}}(\bm{r})$ was chosen to comprise slowly-varying fields, and it is for such variables that the method has most utility, the formal derivations that lead to Eqs.~(\ref{eq:Qa-phi}) and (\ref{eq:Qphi}) also apply to variables that are not slowly-varying. In this case, the full nonlinear and nonlocal equations must be solved to obtain the average fields, a difficult task given the complexity of the equations. For other choices of variables than the slowly-varying densities, there is no compelling reason for generalized thermodynamic relations among conjugate $\bm{\phi}$ fields to be spatially and temporally local, a fundamental assumption of nonequilibrium thermodynamics.  On the other hand, as we shall show in the next section, approximations can be made for slowly-varying fields that lead to tractable hydrodynamic equations for the average fields.

\section{Reaction-diffusion and hydrodynamic equations}\label{sec:RD-hydro}

In this section we derive quantum-classical reaction-diffusion equations for multi-state quantum systems possessing long-lived metastable states, coupled to fluid hydrodynamic equations. For this application, we consider a system of $N$ particles where $N_q$ quantum solute molecules are present in a classical fluid of $N_b$ solvent (bath) molecules $S$. We suppose that quantum particles exist in two long-lived metastable states, denoted by $A$ and $B$, and their corresponding local density operators are given by $\hat{N}_\gamma(\bm{r})$, $\gamma=A,B$. The translational degrees of freedom associated with the centers-of-mass coordinates of the $N_q$ quantum solute molecules will be taken to be classical.~\cite{class-coord} For simplicity, we assume that the internal degrees of freedom of the $N_b$ solvent molecules, if present, may be neglected and only their centers of mass play a role in the dynamics, and all particles are assumed to have a common mass $M$. We then denote the set of classical phase space coordinates $\bm{X}=(R_1, \dots, R_{N}, P_1,\dots,P_{N})=(\bm{R}, \bm{P})$ to be those of the center-of-mass positions and momenta of the solute and solvent molecules. The Hamiltonian of the system is similar to that in Eq.~(\ref{eq:H}), except that the classical Hamiltonian $H_c$ includes the center-of-mass translational energy of the quantum particles, and the interaction potential accounts for interactions between the internal quantum states and the solvent molecules as well as for direct interactions among the quantum states of different quantum particles, $\hat{V}_I(\bm{R})= \hat{V}_{Ib}(\bm{R})+\hat{V}_{Iq}(\bm{R})$, with
\begin{align}
\label{eq:int-pot}
\hat{V}_{I}(\bm{R}) &= \frac{1}{2} \sum_{ (i\neq j) }^{N} \Big[ \left( \Theta_i^q \Theta_j^b + \Theta_i^b \Theta_j^q \right) \hat{V}_{qb}({R}_{ij}) \nonumber \\
& \qquad \qquad +  \Theta_i^q \Theta_j^q \hat{V}_{qq}({R}_{ij}) \Big], \nonumber \\
&= \frac{1}{2} \sum_{ (i\neq j)}^{N}  \hat{V}_{q}({R}_{ij}) \equiv \sum_{i=1}^N \Theta_i^q \hat{V}_{Ii} (\bm{R}_{ci})
\end{align}
where the double sums range over all pairs $(i,j)$ of $i$ and $j$ with $i \neq j$, $R_{ij} = | \bm{R}_j - \bm{R}_i| = | \bm{R}_{ij} |$ is the relative distance between molecules $i$ and $j$, and $\bm{R}_{ci}=\{R_{ij}\mid j=1,\dots,N, i\ne j \} $. The indicator functions $\Theta_x^y$ are unity if $x$ satisfies the property $y$ and zero otherwise.

The slowly varying coarse-grained fields we consider are as follows: the local operators for the metastable chemical species are
\begin{equation*}
\hat{N}_{\gamma}(\bm{r})=\sum_{i=1}^N \Theta_i^q \hat{\mathsf{S}}_i^\gamma(\bm{R}_{bi}) \Delta(\bm{R}_i-\bm{r}),
\end{equation*}
where $\hat{\mathsf{S}}_i^\gamma(\bm{R}_{bi})$ is an operator that specifies when the quantum molecule $i$ is species $\gamma$, and $\bm{R}_{bi}=\{R_{ij}\mid j \in b \} $.  Since this operator satisfies the condition $\sum_\gamma \hat{\mathsf{S}}_i^\gamma(\bm{R}_{bi})=1$, we have $\sum_\gamma \hat{N}_{\gamma}(\bm{r})=\sum_{i=1}^N \Theta_i^q \Delta(\bm{R}_i-\bm{r})=N_q(\bm{r})$, the local density of quantum particles.

The coarse-graining function $\Delta$ vanishes for distances $\ell_{\Delta} \gg \ell_{\rm int}$, where $\ell_{\rm int}$ is the short length scale on which interactions vary rapidly and is introduced to smooth rapid variations that occur in the microscopic density, $\hat{N}_{m\gamma}(\bm{r})=\sum_{i=1}^N \Theta_i^q \hat{\mathsf{S}}_i^\gamma(\bm{R}_{bi}) \delta(\bm{R}_i-\bm{r})$, on $\ell_{\rm int}$ length scales.~\cite{RKO77,RSK24} It satisfies the condition, $\int d\bm{r}' \; \Delta(\bm{r}-\bm{r}')=1$.

The remaining local fluid fields include the number density of the bath molecules,
\begin{equation*}
N_b(\bm{r}) = \sum_{i=1}^{N}\Theta_i^b\Delta(\bm{R}_{i}-\bm{r}),
\end{equation*}
and $N(\bm{r})=N_q(\bm{r})+N_b(\bm{r})$, the total local number density.
The total momentum density of the centers of mass of the solvent and solute molecules is
\begin{equation*}
\bm{g}_N(\bm{r})=
\sum_{i=1}^{N}\bm{P}_{i}\Delta(\bm{R}_{i}-\bm{r}).
\end{equation*}
The Hamiltonian density can be decomposed as $\hat{H}(\bm{r})=\hat{H}_{qc}(\bm{r})+H_c(\bm{r})$, where
\begin{equation*}
\hat{H}_{qc}(\bm{r})= \sum_{i=1}^{N} \Theta_i^q \hat{H}_{qci}(\bm{R})\Delta(\bm{R}_i-\bm{r}),
\end{equation*}
and $\hat{H}_{qci}=\hat{H}_{qi}+ \hat{V}_{Ii}(\bm{R}_{ci})$ for molecule $i$ includes both the quantum Hamiltonian for the particle $\hat{H}_{qi}$ and its interaction $\hat{V}_{Ii}(\bm{R}_{ci})$ with the classical bath particles and other quantum particles, while the classical Hamiltonian density is
\begin{equation*}
H_c(\bm{r})=\sum_{i=1}^{N} \Big( \frac{P_{i}^{2}}{2M}  +U_{ci}\Big) \Delta(\bm{R}_{i}-\bm{r}),
\end{equation*}
and includes interactions among the centers of mass of all particles,
\begin{align*}
U_c &= \frac{1}{2} \sum_{(i\neq j)}^{N}  U_c(R_{ij}) = \sum_{i=1}^N U_{ci}
\end{align*}
where
\begin{align}
\label{eq:classical-potential}
U_c(R_{ij}) &= \Theta_{i}^{b}\Theta_{j}^{b} V_{bb}(R_{ij}) + \Theta_{i}^{q}\Theta_{j}^{q} V^c_{qq}(R_{ij}) \nonumber \\
& +
\left( \Theta_i^q \Theta_j^b + \Theta_i^b \Theta_j^q \right) {V}^c_{qb}(R_{ij}) .
\end{align}
Here, $V^c_{qq}$ and $V^c_{qb}$ denote the (classical) interaction energy between the solute molecular centers of mass and the classical bath particles.

With this choice of variables, the set of local densities is $\widehat{\bm{A}}(\bm{r})$ is~\cite{orthogonal}
\begin{equation*}
\widehat{\bm{A}}(\bm{r})= (\hat{N}_{A}(\bm{r}), \hat{N}_{B}(\bm{r}),N(\bm{r}),
\bm{g}_N(\bm{r}),\hat{H}(\bm{r}) ).
 \end{equation*}
The corresponding $\bm{\phi}$ fields are written as
\begin{eqnarray*}
\bm{\phi}(\bm{r},t)&=&\big( (\beta \tilde{\mu}_{A})(\bm{r},t), (\beta \tilde{\mu}_{B})(\bm{r},t),\\
&& (\beta (\mu_S-\frac{1}{2} m v^2))(\bm{r},t),(\beta \bm{v})(\bm{r},t)),-\beta(\bm{r},t))\big).\nonumber
\end{eqnarray*}
where $\tilde{\mu}_{\gamma } (\bm{r},t)=\mu_{\gamma } (\bm{r},t) - \mu_S (\bm{r},t)$ is the chemical potential of the $\gamma$ species density relative to that of the solvent and, as noted earlier, $\bm{v}(\bm{r},t)$ is the local fluid velocity field, and $\beta(\bm{r},t)$ is related to the inverse local temperature. We have used the notation $(\beta \tilde{\mu})_{A}(\bm{r},t)=\beta(\bm{r},t) \tilde{\mu}_{A}(\bm{r},t)$, with similar expressions for the other fields, to simplify writing the products of fields.

With these expressions for the $\bm{\phi}$ fields we have
\begin{eqnarray}
\label{eq:A*phi}
\widehat{\bm{A}} \ast \bm{\phi}&=& \sum_\gamma\hat{N}_\gamma \ast (\beta \tilde{\mu}_\gamma)   + N \ast (\beta (\mu_S-\frac{1}{2} m v^2)) \nonumber \\
&&+ \bm{g}_N \ast(\beta \bm{v})-\beta \ast \hat{H}.
\end{eqnarray}
Since the numbers of quantum and bath particles can fluctuate due to coupling to the reservoirs, we use the grand canonical local equilibrium operator in Eq.~(\ref{eq:local-equil}) with $\lambda, {\lambda'}\in \{q,b\}$.

The fluxes of $\widehat{\bm{A}}(\bm{r})$ take the general form,
\begin{equation*}
\widehat{\bm{\mathcal{J}}}_{A}(\bm{r})=i\mathcal{L} \widehat{\bm{A}}(\bm{r})=\hat{\bm{\jmath}}_A^{(0)}(\bm{r}) -\partial_{\bm{r}} \cdot \hat{\bm{\jmath}}_A^{(1)}(\bm{r}).
\end{equation*}
For the solute metastable state densities, the flux operators are
$\widehat{{\mathcal{J}}}_{\gamma}=\hat{{j}}_{\gamma}^{(0)} - \partial_{\bm{r}} \cdot \hat{\bm{\jmath}}_{\gamma}^{(1)}$,
with
\begin{eqnarray*}
\hat{{j}}_{\gamma }^{(0)}(\bm{r})&=& \hat{{j}}_{R\gamma }(\bm{r}) = \sum_{i=1}^N \Theta_i^q \big(i\mathcal{L}\hat{\mathsf{S}}_i^\gamma\big) \Delta(\bm{R}_i-\bm{r})
\\
\hat{\bm{\jmath}}_{\gamma}^{(1)}(\bm{r})&=& \hat{{j}}_{\gamma }(\bm{r}) = \sum_{i=1}^N \Theta_i^q \hat{\mathsf{S}}_i^\gamma \frac{\bm{P}_i}{M} \Delta (\bm{R}_i-\bm{r}).
\end{eqnarray*}
We define $ \hat{{j}}_{R\gamma }(\bm{r})=\nu_\gamma  \hat{{j}}_{R }(\bm{r})$ since $\hat{\mathsf{S}}_i^A+\hat{\mathsf{S}}_i^B=1$, with $\nu_\gamma$ the stoichiometric coefficient having values $\nu_A=-1$ and $\nu_B=1$. The fluxes of the local total number, momentum and energy density are
\begin{eqnarray*}
 {{\mathcal{J}}}_{N}&=&-\partial_{\bm{r}} \cdot {\bm{j}}_{N}^{(1)}(\bm{r}) =-\partial_{\bm{r}} \cdot \bm{g}_N(\bm{r})/M \\
 \hat{\bm{\mathcal{J}}}_{g}&=&-\partial_{\bm{r}} \cdot {\hat{\bm{\jmath}}}_{g}^{(1)}(\bm{r}) =- \partial_{\bm{r}} \cdot \hat{\bm{\tau}}(\bm{r})\\
 \widehat{\mathcal{J}}_{e}&=&-\partial_{\bm{r}} \cdot \hat{\bm{\jmath}}_{e}^{(1)}(\bm{r}) =-\partial_{\bm{r}} \cdot \hat{\bm{\jmath}}_e(\bm{r}) ,
\end{eqnarray*}
where the local stress tensor and heat flux have classical and quantum contributions, $\hat{\bm{\tau}}=\bm{\tau}_c + \hat{\bm{\tau}}_q$ and $\bm{j}_e= \bm{j}_{ec} + \hat{\bm{\jmath}}_{eq}$, respectively. Full expressions for these flux contributions are given in Appendix~\ref{app:stress-heat}.

\subsection{Markovian evolution equations for average fields}

The set of nonlocal equations~(\ref{eq:Qa-phi}) take a simpler approximate form for the coarse-grained fields in the set $\widehat{\bm{A}}(\bm{r})$. Since these fields vary on slow hydrodynamics times $\tau_h$, we can utilize the existence of a small parameter $\epsilon \sim \tau_{{\rm mic}}/\tau_h$, where $\tau_{{\rm mic}}$ is a short characteristic microscopic time, to carry out this simplification. In particular, for an operator $\hat{B}$, we have~\cite{RSK24}
\begin{eqnarray*}
\hat{\mathcal{Q}}(t_1)\hat{B} &=& \hat{\mathcal{Q}}(t)\hat{B} - \int_{t}^{t_1} d\tau \, \frac{\delta ( \hat{\mathcal{P}}(\tau)\hat{B})}{\delta \bm{\phi} (\bm{r}_1 ,\tau) } * \dot{\bm{\phi}}(\bm{r}_1,\tau)\nonumber \\
&=& \hat{\mathcal{Q}}(t)\hat{B} + \mathcal{O}(\epsilon) .
\end{eqnarray*}
Consequently, the projectors $\hat{\mathcal{Q}}(t_n)$ in the time-ordered evolution operator $\hat{U}_{Q}(t,t_1)$ can be approximated by $\hat{\mathcal{Q}}(t)$, so that $\hat{U}_{Q}(t,t_1)\approx e^{ i \mathcal{L}\hat{\mathcal{Q}}(t)(t-t_1)}$.

We can also exploit the time scale separation to make a Markovian approximation to the memory term in Eq.~(\ref{eq:Qa-phi}) to obtain
\begin{align*}
\int_0^t & dt_1 \; \big\langle  \hat{\bm{{\mathcal J}}}_{A,t}(\bm{r},t_1,t)\overline{\bm{{\mathcal J}}}_{A,t_1}(\bm{r}_1) ,  \big\rangle^{\rm sym}_{t_1}\ast \bm{\phi}(\bm{r}_1,t_1) \nonumber \\
&\approx  \Big[\int_0^\infty d\tau \; \langle \hat{\bm{\mathcal{J}}}_{A,t}(\bm{r},\tau) \overline{\bm{\mathcal{J}}}_{A,t}(\bm{r}_1) \rangle^{\rm sym}_{t}\Big] \ast \bm{\phi}(\bm{r}_1,t)\nonumber \\
& \equiv \bm{L}_{AA}(\bm{r},\bm{r}_1,t)\ast \bm{\phi}(\bm{r}_1,t),
\end{align*}
with $\hat{\bm{\mathcal{J}}}_{A,t}(\bm{r},\tau)=e^{ i \mathcal{L}\mathcal{Q}(t)\tau}\hat{\bm{\mathcal{J}}}_{A,t}(\bm{r})$. In the Markovian approximation Eq.~(\ref{eq:Qa-phi}) becomes
\begin{eqnarray}\label{eq:Qa-phi-Mark}
\partial_t \bm{a}(\bm{r},t)&=& \langle \hat{\bm{\mathcal{J}}}_{A}(\bm{r}) \rangle_t + \bm{I}(\bm{r},t)  \\
&& \quad -\bm{L}_{AA}(\bm{r},\bm{r}_1,t)\ast \bm{\phi}(\bm{r}_1,t).\nonumber
\end{eqnarray}
When Eq.~(\ref{eq:Qa-phi-Mark}) is written in terms of its components we obtain the set of generalized equations of motion for the average values of the slowly-varying $\bm{\widehat{A}}(\bm{r})$ fields expressed in terms of their conjugate $\bm{\phi}(\bm{r},t)$ fields. These equations are written in full in Appendix~\ref{app:a-eqs-full}.

Once the system of equations is solved for the $\bm{a}(\bm{r},t)$ and their conjugate $\bm{\phi}(\bm{r},t)$ fields, the evolution of the nonequilibrium average $b(\bm{r},t)$ of an arbitrary linear density operator $\hat{B}(\bm{r})$ at long times can be expressed in terms of local density averages,
\begin{align*}
\partial_t b(\bm{r},t) &= \langle \hat{\cal{J}}_B (\bm{r},t) \rangle_t + \bm{I}_B(\bm{r},t)  \\
& \quad -\bm{L}_{BA}(\bm{r},\bm{r}_1,t)\ast \bm{\phi}(\bm{r}_1,t) ,\nonumber
\end{align*}
where $\bm{I}_B(\bm{r},t)$ and $\bm{L}_{BA}(\bm{r},\bm{r}_1,t)$ are obtained by replacing $i{\cal L} \widehat{\bm{A}}(\bm{r})$ by $i{\cal L} B(\bm{r})$ in $\bm{I}(\bm{r},t)$ and $\hat{\cal{\bm{J}}}_{A,t}(\bm{r},t)$.

It is often useful to introduce two additional approximations to make the evaluation of the correlation functions more tractable. First, since the $\bm{\phi}$ fields and their gradients are slowly varying in space, we can expand $\bm{\phi}(\bm{r}_1,t)$ in a Taylor series about $\bm{r}$ since the fluxes in the local equilibrium correlation function are significantly correlated only when $|\bm{r}-\bm{r}_1|$ is small. This approximation allows us to write $\bm{L}_{AA}(\bm{r},\bm{r}_1,t)\ast \bm{\phi}(\bm{r}_1,t) \approx \bm{L}_{AA}(\bm{r},t)\cdot \bm{\phi}(\bm{r},t)$, where
\begin{equation*}
\bm{L}_{AA}(\bm{r},t)=\int_0^\infty d\tau \; \langle \hat{\bm{\mathcal{J}}}_{A,t}(\bm{r},\tau) \overline{\bm{\mathcal{J}}}_{A,t} \rangle^{\rm sym}_{t},
\end{equation*}
with $\overline{\bm{\mathcal{J}}}_{A,t} \equiv \int d \bm{r}_1 \; \overline{\bm{\mathcal{J}}}_{A,t}(\bm{r}_1)$.
Second, local equilibrium averages may be approximated by homogeneous ensemble averages~\cite{KO88} where operators or variables $\hat{B}(\bm{r})$ in the correlation functions are replaced by their volume averages so that the transport properties in the nonequilibrium ensemble depend on space and time only through the $\bm{\phi}(\bm{r},t)$ fields. With this approximation, we can write
\begin{equation*}
\bm{L}_{AA,H}(\bm{r},t)=\frac{1}{V}\int_0^\infty d\tau \; \langle \hat{\bm{\mathcal{J}}}_{A,t}(\tau) \overline{\bm{\mathcal{J}}}_{A,t} \rangle^{\rm sym}_H,
\end{equation*}
where the subscript $H$ on the angular brackets denotes an average in the homogeneous ensemble where the local density $\hat{\rho}_L(t)$, Eq.~(\ref{eq:local-equil}), is replaced by
\begin{equation}
\label{eq:local_eq-homo}
 \hat{\rho}_{H}(\bm{r},t) = \prod_{\lambda}(N_{\lambda} ! h^{3N_{\lambda}})^{-1} e^{\widehat{\bm{A}} \cdot \bm{\phi}_{A}(\bm{r},t)}/{\cal Z}_H(t) ,
\end{equation}
where 
\begin{align*}
    \mathcal{Z}_H(t)={\rm Tr}\left[ \prod_{\lambda'}(N_{\lambda'} ! h^{3N_{\lambda'}})^{-1} e^{\widehat{\bm{A}} \cdot \bm{\phi}(\bm{r},t)}\right] .
\end{align*}
In the homogeneous ensemble Eq.~(\ref{eq:A*phi}) reads
\begin{align}
\widehat{\bm{A}} \cdot &\bm{\phi}(\bm{r},t) =\sum_\gamma\hat{N}_\gamma  (\beta \tilde{\mu}_\gamma)(\bm{r},t) + \bm{g}_N \cdot(\beta \bm{v})(\bm{r},t)    \nonumber \\
&+ N  (\beta (\mu_S-\frac{1}{2} m v^2))(\bm{r},t) -\beta(\bm{r},t)  \hat{H}. \label{eq:A*phi-homo}
\end{align}
As noted above, in this expression the dynamical variables and operators no longer depend on space, and spatial dependence resides only in the $\bm{\phi}$ fields. It follows from the form of $\hat{\rho}_H(\bm{r},t)$ that the dependence of homogeneous local equilibrium correlation functions on the local conjugate fields $\bm{\phi}(\bm{r},t)$ is the same\cite{KO88} as the dependence of equilibrium correlation functions on the equilibrium values of $\bm{\phi}$. 
 Furthermore, standard thermodynamic relations, such as the Gibbs-Duhem equation, hold locally at the field point $\bm{r}$.
 
Lastly, the initial condition term $\bm{I}(\bm{r},t)$ will decay on a microscopic time scale, $\tau_{\rm mic}$, and for times $t > \tau_{\rm mic}$ it may be neglected. Applying these approximations, the evolution equation is,
\begin{equation*}
\partial_t \bm{a}(\bm{r},t)= \langle \hat{\bm{\mathcal{J}}}_{A}(\bm{r}) \rangle_t -\bm{L}_{AA,H}(\bm{r},t)\cdot \bm{\phi}(\bm{r},t),
\end{equation*}
and this is the form that will be considered in what follows. To avoid excessive notation, the subscript $H$ will be discarded in expressions for dissipative coefficients $\bm{L}_{AA}$, and it should be assumed that the flux correlation functions are evaluated in the homogeneous ensemble.

Any of the approximations above can be relaxed if the conditions for their validity are violated when applying this formalism to a physical system, albeit at the cost of dealing with more complicated equations.

\section{Reaction coordinates and quantum representation}\label{sec:Rcoord}

The results obtained in the previous sections apply to any choice of species variables to specify the quantum operators or functions that define the metastable $A$ and $B$ species. Also, any convenient basis can be used to represent the quantum operators. Here we consider some choices for species variables and basis sets to illustrate the use of the general formalism.

The choice of chemical species variables depends on the reaction under consideration. In some circumstances, it may be convenient to consider a representation based on the eigenfunctions of the quantum Hamiltonian $\hat{H}_q=\sum_{i=1}^N \Theta_i^q \hat{H}_{qi}$: $\hat{H}_{qi} \mid ki \rangle =\epsilon_{ki} \mid ki \rangle$, where $ki \in\{ k_1,k_2, \dots, k_s\}$ for an s-state quantum molecule. The interactions may be such that these quantum states can be partitioned into two weakly-coupled submanifolds $\{s_\gamma\}$ corresponding to $A$ and $B$, specified by the operators $\hat{\mathsf{S}}^\gamma_i = \sum_{ki \in \{s_\gamma\}} \mid ki \rangle \langle ki \mid$.

In other applications, the species variables may be chosen to depend only on the environmental coordinates. This is the case for proton or electron transfer reactions controlled by the solvent polarization reaction coordinate, as mentioned in the Introduction. In such circumstances, the species variables are classical and depend on the coordinates of the bath: $\mathsf{S}^\gamma_i(\bm{R}_{bi})$. In other physical situations, chemical species may be characterized by quantum operators that are functions of the environmental coordinates, as is the case for a species specification in terms of adiabatic states.

Here we suppose that the reaction is governed by a general scalar reaction coordinate $\xi_i(\bm{R}_{bi})$, which is a function of the bath coordinates relative to the position of the quantum particle $i$, which can be used to define a hypersurface $\xi^\ddagger$ that partitions the configuration space into two metastable regions. The corresponding species variable may be chosen to be $\mathsf{S}^\gamma_i(\bm{R}_{bi})=\theta(\nu_\gamma(\xi_i(\bm{R}_{bi})-\xi^\ddagger))$, where $\theta(x)$ is the Heaviside function.

To investigate various aspects of the general formulation in a simple context, we assume that the thermal conductivity is sufficiently large so that the temperature is uniform and constant, $\beta(\bm{r},t)=\beta$. For strongly exothermic or endothermic reactions, this condition can be relaxed. Chemostats that control the chemical potentials of the solute species are present, and we take the solvent chemical potential to be uniform. As in isotropic systems close to equilibrium, the dominant dynamical correlations are assumed to be due to correlations between fluxes of the same tensorial character~\cite{GM62}, and we retain only these dissipative coefficients, although the presence of constraints may lead to violations of these symmetries. Although dissipative coefficients that couple reaction to the longitudinal component of the velocity field remain in this approximation, this coupling is often small and vanishes for the incompressible fluids ($\partial_{\bm{r}} \cdot \bm{v}=0$) we consider here. As observed earlier, these restrictions can be relaxed by using the full set of generalized hydrodynamic equations given in Appendix~\ref{app:a-eqs-full}, but with these approximations, the equations take the simpler form,
\begin{eqnarray}
\partial_t n_\gamma &=&- \bm{v} \cdot \partial_{\bm{r}}  n_\gamma  +\nu_\gamma \langle j_R \rangle_t  - \nu_\gamma \beta {L}_{ R}   {\mathcal{A}} \label{eq:av-ngamma} \\
&&   +  \partial_{\bm{r}} \cdot  \beta \bm{L}_{\gamma \gamma'} \cdot \partial_{\bm{r}} \widetilde{\mu}_{\gamma'}
\nonumber \\
\partial_t \rho &=&  - \bm{v} \cdot \partial_{\bm{r}} \rho \nonumber \\
\rho \partial_t \bm{v}  &=&   - \partial_{\bm{r}}\cdot  \bm{P}_h   +\partial_{\bm{r}} \cdot \beta \bm{L}_{v v} :  \partial_{\bm{r}} \bm{v}
\label{eq:av-v}.
\end{eqnarray}
The summation convention on repeated indices is used here and below.

Sometimes it is useful to rewrite this set of equations in a more standard form by evaluating the average fluxes and expressing the dissipative $\bm{L}$ coefficients in terms of usual transport coefficients. The full expression for the reactive flux is
\begin{equation*}
  j_{R}(\bm{r})=\sum_{i=1}^N \Theta_i^q \frac{\bm{P}}{M} \cdot ( \partial_{\bm{R}} \mathsf{S}^B_i(\bm{R}_{bi})) \Delta(\bm{R}_{i}-\bm{r}),
\end{equation*}
and its average in the nonequilibrium ensemble is
\begin{eqnarray*}
&& \langle j_R(\bm{r}) \rangle_t=\bm{\mathcal{E}}_R(\bm{r},t) \cdot  \bm{v}(\bm{r} , t) \\
&& \qquad = \Big\langle \sum_{i=1}^N \Theta_i^q ( \partial_{\bm{R}}\mathsf{S}^B_i(\bm{R}_{bi})) \Delta(\bm{R}_{i}-\bm{r})\Big\rangle_t \cdot \bm{v}(\bm{r} , t),\nonumber
\end{eqnarray*}
which is localized on the reaction hypersurface since $\partial_{\bm{R}} \mathsf{S}^B_i=( \partial_{\bm{R}} \xi_i )\delta(\xi_i -\xi^\ddagger)$. The reactive coefficient ${L}_{ R}$ is defined by
\begin{equation}
\label{eq:LR}
\beta {L}_{ R}(\bm{r},t)=\frac{\beta}{V}\int_0^\infty d \tau \;\big\langle  \widetilde{j}_{R,t}(\tau) \overline{j}_{R,t}   \big\rangle^{\rm sym}_H,
\end{equation}
where the spatially-integrated fluxes that enter the homogeneous ensemble are $ {j}_{R,t}=\mathcal{Q}(t)j_R$. This is a generalization of equilibrium reactive flux correlation function expressions~\cite{yamamoto60}; it is space and time-dependent due to the nonequilibrium homogeneous ensemble average.

The diffusion flux is
\begin{equation*}
  {\bm{j}}_{\gamma, t}= \mathcal{Q}(t)\sum_{i=1}^N \Theta_i^q \mathsf{S}^\gamma_i(\bm{R}_{bi})   \frac{\bm{P}_i}{M},
\end{equation*}
and the corresponding $A$ and $B$ diffusion dissipative coefficients are
\begin{eqnarray*}
\beta \bm{L}_{\gamma \gamma' }(\bm{r},t)&=&\frac{\beta}{V}\int_0^\infty d \tau \;\big\langle  \widetilde{\bm{j}}_{\gamma,t}(\tau) \overline{\bm{j}}_{\gamma',t}   \big\rangle^{\rm sym}_H\\
&\equiv&\beta \bm{D}_{\gamma \gamma'} (\bm{r},t) n_\gamma(\bm{r},t) ,
\end{eqnarray*}
where the last line defines the local diffusion tensor.

We let $ \langle{}\bm{\tau}^{+}(\bm{r})\rangle_{t}=\bm{P}_h(\bm{r},t) $ define the local fluid hydrostatic pressure tensor, while the viscous dissipative coefficient is a rank-4 tensor,
\begin{eqnarray*}
\beta\bm{L}_{vv}(\bm{r},t )&=&\frac{\beta}{V}\int_0^\infty d\tau \; \langle \hat{\bm{\tau}}_{t}(\tau) \overline{\bm{\tau}}_{t} \rangle^{\rm sym}_H .
\end{eqnarray*}
As noted above, all of these transport coefficients depend on the spatial field point $\bm{r}$ and the time $t$ because the homogeneous averages depend on the conjugate fields $\bm{\phi}(\bm{r},t)$.

We can use the relation in Eq.~(\ref{eq:adot-phidot}), along with the expression $\langle \widetilde{N}_\gamma(\bm{r}) \overline{N}_\gamma(\bm{r}_1) \rangle_t \approx n_\gamma(\bm{r},t) \delta_{\gamma \gamma'} \delta(\bm{r}-\bm{r}_1)$ for a dilute solution to find
\begin{equation}
\label{eq:gammadot-mu-gammahidot}
\partial_t n_\gamma(\bm{r},t)=n_\gamma(\bm{r},t)\beta  \partial_t{\mu}_\gamma(\bm{r},t).
\end{equation}
Taking the initial values of the fields to be uniform, this equation admits a solution, ${\mu}_\gamma(\bm{r},t)=\mu_\gamma^0 +\beta^{-1} \ln (n_\gamma(\bm{r},t)/n_0)$. If this result is used in Eqs.~(\ref{eq:av-ngamma})-(\ref{eq:av-v}), one has a closed set of coupled equations to solve self-consistently for the $n_\gamma(\bm{r},t)$ and $\bm{v}(\bm{r},t)$ fields.
Alternatively, one can use Eq.~(\ref{eq:gammadot-mu-gammahidot})  in Eq.~(\ref{eq:Qa-phi}) to obtain coupled equations for the $\mu_\gamma(\bm{r},t)$ and $\bm{v}(\bm{r},t)$ fields.

\subsection{Evaluation of the dissipative coefficients}
The correlation function expressions for the dissipative coefficients can be evaluated in any basis. Because the physical system under investigation comprises identifiable quantum solute species in a solvent, the interactions among the solute species must not be so strong that these species lose their identity. In this circumstance, a molecular-based adiabatic basis for the configuration-space reaction coordinate considered here is especially convenient. In this basis the matrix elements of direct interaction potentials $\hat{V}_{Iq}$ between the quantum particles will have off-diagonal components, as will the local density operator. Here, for simplicity, we assume that the $N_q$ quantum solute molecules are dilutely dispersed in the classical fluid, so that $N_q \ll N_b$ and we can neglect the off-diagonal direct interactions between quantum particles that lead to quantum transitions.

In this case the Hamiltonian $\hat{H}_{qc}$ can be decomposed into the sum of single noninteracting quantum particle Hamiltonians, $\hat{H}_{qci}$. The corresponding adiabatic states for particle $i$ are defined by
\begin{equation*}
\hat{H}_{qci}(\bm{R}_{bi})\mid {\ell}i; \bm{R}_{bi} \rangle  = \epsilon_{{\ell}i}(\bm{R}_{bi}) \mid {\ell}i; \bm{R}_{bi} \rangle ,
\end{equation*}
with $\ell i \in \{ \ell_1, \ell_2, \dots, \ell_s \}$, $\forall i$, for a particle with $s$ quantum states. The $N_q$ quantum particles are distinguishable by their classical center-of-mass coordinates, and the adiabatic states for the entire system can be written as $\mid \bm{\ell}; \bm{R} \rangle =\prod
_{i=1}^N \Theta_i^q \mid {\ell}i; \bm{R}_{bi} \rangle$, so that
\begin{equation*}
\hat{H}_{qc}(\bm{R})\mid \bm{\ell}; \bm{R} \rangle  = \epsilon_{\bm{\ell}}(\bm{R}) \mid \bm{\ell}; \bm{R} \rangle ,
\end{equation*}
with $\epsilon_{\bm{\ell}}(\bm{R})= \sum_{i=1}^N \Theta_i^q \epsilon_{{\ell}i}(\bm{R}_{bi})$.

For the situations considered here, where $\beta$ and $\mu_S$ are constant, we can write $\widehat{\bm{A}} \cdot \bm{\phi}(\bm{r},t)$ in Eq.~(\ref{eq:A*phi-homo}) as
\begin{eqnarray}
\label{eq:A*phi-homo-2}
&&\widehat{\bm{A}} \cdot \bm{\phi}(\bm{r},t)=
-\beta \Big( \hat{H}^+ - N_b  \mu_S - \sum_\gamma  {N}_\gamma  \mu_\gamma(\bm{r},t) \Big)  \\
&&\qquad = -\beta \Big( \hat{H}^+ - N_b  \mu_S - {N}_q  \mu_q(\bm{r},t)  + {N}_r  \mathcal{A}(\bm{r},t)\Big), \nonumber
\end{eqnarray}
where $\hat{H}^+$ is $\hat{H}$ with $\bm{P}_i \to \bm{P}^+_i=\bm{P}_i-M\bm{v}(\bm{R}_i)$, $N_q=N_A + N_B$ defined earlier, $\mu_q=(\mu_A+\mu_B)/2$ and $N_r=(N_A - N_B)/2$. In the second line of Eq.~(\ref{eq:A*phi-homo-2}) we expressed $\widehat{\bm{A}} \cdot \bm{\phi}$ in terms of the total density of quantum particles $N_q$, controlled by $\mu_q$, which can often be taken to be constant since the reaction interconverts species $A$ and $B$ without changing their total number, and the difference density $N_r$ controlled by the chemical affinity $\mathcal{A}(\bm{r},t)= \mu_B(\bm{r},t)-\mu_A(\bm{r},t)$ is the chemical affinity.  We then have the following matrix elements:
\begin{eqnarray*}
{N}_\gamma^{\bm{\ell}' \bm{\ell}}&=&\sum_{i=1}^N \Theta_i^q \mathsf{S}^\gamma_i(\bm{R}_{bi})  \delta_{\bm{\ell}' \bm{\ell}} \equiv  {N}_\gamma \delta_{\bm{\ell}' \bm{\ell}} \\
{H}^{\bm{\ell}' \bm{\ell}}_{qc} &=& \sum_{i=1}^N \Theta_i^q \epsilon_{\ell i} (\bm{R}_{bi}) \delta_{\bm{\ell}' \bm{\ell}} = \epsilon_{\bm{\ell}}(\bm{R}) \delta_{\bm{\ell}' \bm{\ell}},
\end{eqnarray*}
with $\delta_{\bm{\ell}' \bm{\ell}}=\prod_{j=1}^N \Theta_j^q \delta_{{\ell}'j {\ell}j}$. Using these results, the matrix elements of $\widehat{\bm{A}} \cdot \bm{\phi}$ are
\begin{equation*}
{\bm{A}}^{\bm{\ell}' \bm{\ell}} \ast \bm{\phi}=   -\beta \Big[  \mathcal{H}_{\bm{\ell}}^+  + N_r \mathcal{A}(\bm{r},t) \Big]  \delta_{\bm{\ell}' \bm{\ell}},\nonumber
\end{equation*}
with $\mathcal{H}_{\bm{\ell}}^+ = H_c^+ +\epsilon_{\bm{\ell}} -N_b \mu_S -N_q \mu_q$,
so that
\begin{eqnarray} \label{eq:rhoH-ell}
\rho_{H}^{\bm{\ell}' \bm{\ell}}(\bm{X} ,\bm{r}, t)&=&\Big(\prod_{\lambda}(N_{\lambda} ! h^{3N_{\lambda}})^{-1} e^{-\beta (\mathcal{H}_{\bm{\ell}}^+ +N_r \mathcal{A}(\bm{r},t)}/\mathcal{Z} \Big)\delta_{\bm{\ell}' \bm{\ell}} \nonumber \\
&=&\rho_{H}^{\bm{\ell}} (\bm{r},t) \delta_{\bm{\ell}' \bm{\ell}}, 
\end{eqnarray}
where $\lambda =b,q,r$. Note that $N_q$ is independent of the configuration space coordinates because the sum of the solute species variables is unity, but that is not the case for $N_r$ since it depends on the difference of these variables. Because $\rho_{H}^{\bm{\ell}' \bm{\ell}}$ is diagonal in the adiabatic basis the computation of average values simplifies,
\begin{equation*}
\langle \hat{B}(\bm{X}) \rangle_{H} (\bm{r},t) = {\rm Tr} \big[{B^{\bm{\ell} \bm{\ell}}}(\bm{X}) \rho_{H}^{\bm{\ell} \bm{\ell}} (\bm{X}, \bm{r},t) \big],
\end{equation*}
where ${\rm tr}= \sum_{\bm{\ell}}$ in the definition of ${\rm Tr}$ in Eq.~(\ref{eq:Trace}).

As an example of the calculation of dissipative coefficients in the adiabatic basis, we can write the correlation function for $L_R$ in Eq.~(\ref{eq:LR}) as
\begin{equation*}
{L}_{ R}(\bm{r},t)=
\frac{1}{V}
\int_0^\infty d \tau {\rm Tr} \Big[ \Big(\big(e^{i \mathcal{L} \mathcal{Q}(t) \tau}\big)_{\bm{\ell} \bm{\ell}} {j}_{R,t}\Big) {j}_{R,t}  \rho_{H}^{\bm{\ell}} (\bm{r},t) \Big],\nonumber
\end{equation*}
where the matrix elements of the reactive fluxes are diagonal so that $j_{R,t}^{\bm{\ell}' \bm{\ell}}= j_{R,t} \delta_{\bm{\ell}' \bm{\ell}}$ (and $j_{\gamma,t}^{\bm{\ell}' \bm{\ell}}= j_{\gamma,t} \delta_{\bm{\ell}' \bm{\ell}}$ for the diffusion correlation function). For our choice of species variable $j_{R,t}$ is classical and commutes with the density operator so $\overline{j}_{R,t} = j_{R,t}$. Although $j_{R,t}$ is classical, nonadiabatic effects enter ${L}_{ R}(\bm{r},t)$ since evolution governed by the quantum-classical Liouville operator is not diagonal in this basis. Furthermore, the statistical average over the local density operator involves a sum over all quantum states.

The methods used to compute reactive flux correlation functions can be applied here as well. In particular, one can exploit the time scale separation between the chemical and microscopic relaxation times to remove the projected dynamics when a plateau exists for times $t_{\rm mic}\ll t \ll t_{\rm chem}$.~\cite{C78,0chap-kapral98} The evolution under quantum-classical Liouville dynamics can be simulated using various methods, and rare event sampling can be used to evaluate the correlation function.~\cite{T2010} The use of the auxiliary local density operator is an important feature of the present formulation that allows one to bypass sampling from the full quantum or quantum-classical density matrix.

Similar expressions can be written for the other dissipative coefficients, and the matrix elements of the other fluxes are given in Appendix~\ref{app:stress-heat}.

\subsection{Nonequilibrium steady states}
Next, we consider steady states for systems in mechanical equilibrium where $\rho \partial_t \bm{v}=0$, and further suppose that the velocity field is sufficiently small that its effects can be neglected.~\cite{GM62} In this case we are left with the reaction-diffusion equations,
\begin{equation*}
\partial_t n_\gamma =  - \nu_\gamma \beta {L}_R{\mathcal{A}} +  \partial_{\bm{r}} \cdot  \beta \bm{L}_{\gamma \gamma} \cdot \partial_{\bm{r}} \mu_{\gamma},
\end{equation*}
where we have neglected the cross-coupling of diffusion fluxes in $\bm{L}_{\gamma \gamma}$ since the solution is dilute.  Making use of Eq.~(\ref{eq:gammadot-mu-gammahidot}) and the expression for the chemical potential in terms of the local species density, we can write a closed set of equations for the chemical potentials needed to satisfy the constraints:
\begin{equation*}
 n_0 \partial_t \mu_\gamma = e^{-\beta (\mu_\gamma -\mu^0_\gamma)}\big[  - \nu_\gamma {L}_R{\mathcal{A}} +  \partial_{\bm{r}} \cdot \bm{L}_{\gamma \gamma} \cdot \partial_{\bm{r}} \mu_{\gamma}\big] ,
\end{equation*}
where $\mu_\gamma^0$ is the (uniform) chemical potential of species $\gamma$ in equilibrium.  In addition to the manifest nonlinearity, the dissipative coefficients are also functions of the nonlocal chemical potentials.

In place of such a calculation, we can instead consider the somewhat simpler reaction-diffusion equations expressed in terms of the local species densities,
\begin{equation*}
\partial_t n_\gamma = - \nu_\gamma \beta {L}_{R}   \mathcal{A} +  \partial_{\bm{r}} \cdot \bm{D}_\gamma \cdot \partial_{\bm{r}} n_{\gamma}.
\end{equation*}
The reactive term can be written in mass-action form~\cite{RSK24} using the expressions for the chemical potentials derived earlier. To do this we first write the affinity as
\begin{eqnarray*}
&&\beta  \mathcal{A}(\bm{r},t)= \beta(\mu_B-\mu_A) =\beta( \mu_B^0-\mu_A^0)+ \ln \frac{n_B(\bm{r},t)}{n_A(\bm{r},t)} \nonumber \\
  &&\quad = \ln \frac{n_B(\bm{r},t)n_A^{\rm eq}}{n_A(\bm{r},t)n_B^{\rm eq}} \approx \Big(\frac{n_B(\bm{r},t)}{n_B^{\rm eq}}  -\frac{n_A(\bm{r},t)}{n_A^{\rm eq}}  \Big),
\end{eqnarray*}
since $\mathcal{A}=0$ at equilibrium. The last approximate equality follows if the system is close to equilibrium. We may then write
\begin{eqnarray*}
\nu_\gamma \beta {L}_{R}   \mathcal{A}&=& \nu_\gamma \Big(\frac{ {L}_{R}}{n_B^{\rm eq}} n_B(\bm{r},t)- (\frac{ {L}_{R}}{n_A^{\rm eq}} n_A(\bm{r},t)   \Big) \nonumber \\
&\equiv&  \nu_\gamma \big( k_r n_B(\bm{r},t)- k_f n_A(\bm{r},t)   \big).
\end{eqnarray*}
The forward and reverse rate coefficients, $k_{f,r}= k_{f,r}[n_\gamma(\bm{r},t)]$, are functionals of the local densities given the nonequilibrium averages in $L_R$. Because the reaction rate and diffusion coefficients depend on the species densities, we have coupled nonlinear equations,
\begin{eqnarray*}
  \partial_t n_A &=&  k_r n_B- k_f n_A  +  \partial_{\bm{r}} \cdot \bm{D}_A \cdot \partial_{\bm{r}} n_A   \\
  \partial_t n_B &=& - k_r n_B+ k_f n_A  +  \partial_{\bm{r}} \cdot \bm{D}_B \cdot \partial_{\bm{r}} n_B. \nonumber
\end{eqnarray*}
These equations also admit nonequilibrium steady-state solutions $n_\gamma^{\rm ss}(\bm{r})$ that satisfy the differential equations,
\begin{eqnarray}
\label{eq:ss-RD}
&&  \partial_{\bm{r}} \cdot \bm{D}_A \cdot \partial_{\bm{r}} n_A^{\rm ss} +  k_r n_B^{\rm ss}- k_f n_A^{\rm ss} =   0    \\
&&  \partial_{\bm{r}} \cdot \bm{D}_B \cdot \partial_{\bm{r}} n_B^{\rm ss}  - k_r n_B^{\rm ss}+ k_f n_A^{\rm ss} =  0  . \nonumber
\end{eqnarray}
The inhomogeneous steady states depend on the boundary conditions that fix the chemical potentials or densities of the reactive species.

In the linear regime, the transport coefficients can be evaluated using equilibrium averages to obtain the first approximations to the steady-state densities. Then, if one considers a system with two chemostats separated by a distance $z_0$ that fix the concentrations of the two chemical species at $z_\pm=\pm z_0/2$, the one-dimensional version of Eq.~(\ref{eq:ss-RD}) can be solved to yield the nonequilibrium steady state densities. For equal diffusion coefficients, $D_A=D_B=D$, the coupled equations can be solved using the variables $\psi(z)=k_f n_A(z) - k_r n_B(z)$  and $n(z)=n_A(z)+n_B(z)=n_0$ (since reaction conserves the total number of quantum particles), to give
\begin{eqnarray*}
n^{\rm ss}_A(z)&=&\big(k_r n_0 +\psi(z) \big)/(k_f +k_r)\nonumber \\
n^{\rm ss}_B(z)&=&\big(k_f n_0 -\psi(z)\big)/(k_f +k_r),
\end{eqnarray*}
where
\begin{eqnarray*}
\psi(z)&=& 2 k_f n_0 \frac{\cosh (\kappa z_0/2)}{\sinh \kappa z_0} \sinh \kappa z\nonumber \\
&&-\frac{(k_f + k_r)}{\sinh \kappa z_0} \Big( n_B(z_0/2) \sinh \kappa(z+z_0/2)\nonumber \\
&& - n_B(-z_0/2) \sinh \kappa(z-z_0/2)  \Big),\nonumber
\end{eqnarray*}
where $\kappa=((k_f +k_r)/D)^{1/2}$ is the inverse screening length. Chemical reactions screen the diffusive decay of concentration inhomogeneities: Since the chemical relaxation time is $\tau_{\rm chem}= (k_f +k_r)^{-1}$, the screening length corresponds to the average diffusive distance in a chemical relaxation time. Consequently, the dimensionless factor $\kappa z_0$ is important in determining the behavior on the scale of the separation between the two reservoirs. The rate coefficients enter the species densities as ratios that can be expressed in terms of the equilibrium constant for the reaction: $k_f/(k_f +k_r)= 1/(1+K_{\rm eq}^{-1})$ and $k_r/(k_f +k_r)= 1/(K_{\rm eq} +1)$. Thus, the dimensionless screening length $\kappa z_0$ and equilibrium constant $K_{\rm eq}=k_f/k_r= n_B^{\rm eq}/n_A^{\rm eq}$ determine the forms taken by the steady-state densities.

The spatially-dependent steady-state affinity can also be expressed in terms of the steady-state densities as
\begin{eqnarray*}
\beta \mathcal{A}^{\rm ss}(z)&=& \beta(\mu_B^0 -\mu_A^0) + \ln \Big(\frac{n_B^{\rm ss}(z)}{n_A^{\rm ss}(z) }  \Big)\\
&=& \ln \Big(\frac{n_B^{\rm ss}(z)}{(1-n_B^{\rm ss}(z)) K_{\rm eq}}  \Big).
\end{eqnarray*}

\begin{figure}[htbp]
\centering
\resizebox{0.9\columnwidth}{!}{
      \includegraphics{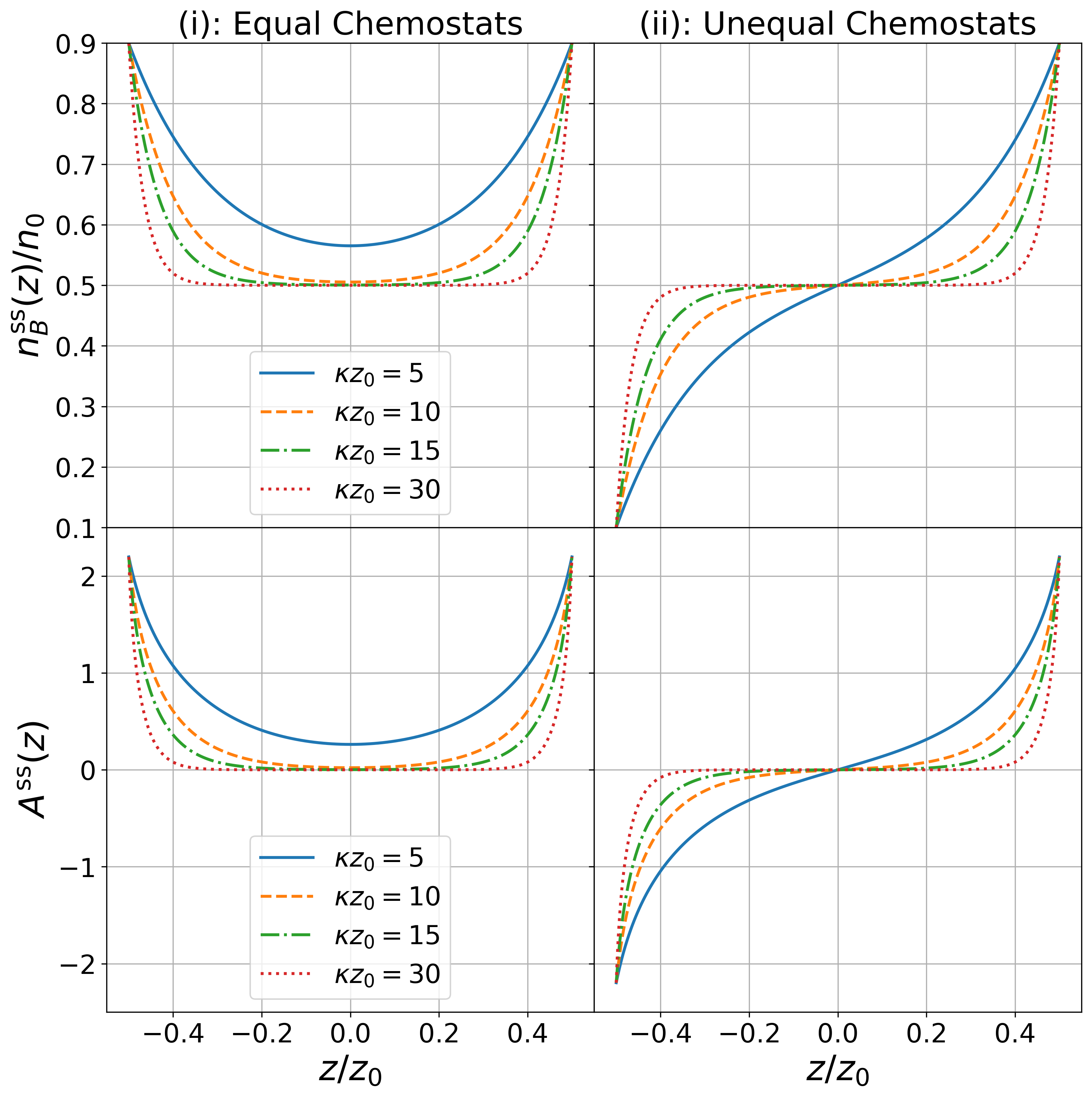}}
        \caption{Plots of the steady-state density $n_B^{\rm ss}(z)/n_0$ (upper two panels) and chemical affinity $\beta \mathcal{A}^{\rm ss}(z)$ (lower two panels) as function of $z$ for several values of $\kappa z_0=5, 10, 15, 30$ indicated in the figures. Results are shown for two chemostat concentration sets (i) and (ii) given in the text: (i) (left panels) and (ii) right panels. The equilibrium constant is $K_{\rm eq}=1$ for these plots.
    } \label{fig:4-panel}
\end{figure}
The steady-state density $n^{\rm ss}_B(z)$ and the chemical affinity $\mathcal{A}^{\rm ss}(z)$ are plotted in Fig.~\ref{fig:4-panel} for two specific chemostat concentrations: (i) the two chemostats have the same concentrations, $(n_A(\pm z_0/2), n_B(\pm z_0/2))/n_0=(0.1,0.9)$ and (ii) the two chemostats have different concentrations, $(n_A(z_0/2), n_B(z_0/2))/n_0=(0.1,0.9)$ and $(n_A(- z_0/2), n_B(- z_0/2))/n_0=(0.9,0.1)$, and several values of $\kappa z_0$. Some features of the plots are common to both chemostat configurations. For small $\kappa z_0$ values $n^{\rm ss}_B(z)$ (and $n^{\rm ss}_A(z)$) vary strongly with $z$ over the entire domain $z_0$ between the chemostats. By contrast, for large values of $\kappa z_0$ significant variations occur only in the regions near the chemostats, while there is a large central region where $n^{\rm ss}_B(z) \approx n^{\rm eq}_B = n_0/2$. The steady-state affinity shows similar trends and in the central region between the chemostats, one has $\mathcal{A}^{\rm ss}(z) \approx 0 $ for large $\kappa z_0$, signaling that near equilibrium conditions exist there.

The inverse screening length can be written as $\kappa =(D \tau_{\rm chem})^{-1/2}$ and for a fixed value of the diffusion coefficient its value is controlled by the chemical relaxation time. Thus, for slow reactions where $\tau_{\rm chem}$ is large, $\kappa$ is small, and small values of $\kappa z_0$ imply that reactions are unable to establish chemical equilibrium on lengths $z_0$, the distance between the chemostats. In the opposite limit where reactions are fast and $\kappa$ is large, chemical equilibrium is established on length scales much shorter than $z_0$, so that large deviations from equilibrium are confined to distances $\ell_s =\kappa^{-1}$ near the chemostats.

The steady-state affinity can be used compute the next approximation to the dissipative reactive coefficient,
\begin{eqnarray*}
\beta {L}^{\rm ss}_{ R}(z)&=&\frac{\beta}{V}\int_0^\infty d \tau \;{\rm Tr} \Big[ \big(\big(e^{i \mathcal{L} \mathcal{Q}(t) \tau}\big)_{\bm{\ell} \bm{\ell}} {j}_{R,t}\big) {j}_{R,t} \rho_{H,ss}^{\bm{\ell}} (z)\Big],\nonumber 
\end{eqnarray*}
where $\rho_{H,ss}^{\bm{\ell}} (z)$ is given by Eq.~(\ref{eq:rhoH-ell}) with the chemical affinity replaced by its steady state value, $\mathcal{A}^{\rm ss}(z)$~\cite{nonlinear-rate}. Since the method is constructed to give the exact nonequilibrium average species densities, the nonlinear effects due to a spatially dependent affinity enter through the dissipative coefficients in the equations of motion, which can be used to investigate the evolution of deviations from steady states. Because $\mathcal{A}^{\rm ss}(z)$ exhibits strong variations with $z$, especially near the chemostats, there are significant $z$-dependent effects on the effective reaction rate coefficients due to the nonequilibrium nature of the environment. In the homogeneous ensemble approximation in Eq.~(\ref{eq:local_eq-homo}), the dependence of the transport coefficients on the local internal energy and the number density is the same as it is in equilibrium\cite{KO88}, and this feature can be exploited to obtain approximate solutions to investigate nonlinear effects far from equilibrium.

\section{Conclusion} \label{sec:conc}
It is useful to summarize several features of the treatment of open quantum-classical systems described in this work. The focus of the study was on the derivation of equations of motion for a set of operators and functions that vary slowly compared to microscopic time scales. The choice of variables depends on the system and is usually selected on physical grounds, but, generally, these variables may be associated with conserved or nearly conserved quantities. In the application studied in this paper, they include the conserved mass, momentum, and energy fields, as well as nearly-conserved reactive species densities corresponding to long-lived metastable states. The densities of conserved quantities must be included in the set of slow variables to ensure a time-scale separation between the evolution of these fields and other microscopic variables.  Their inclusion is necessary to allow a Markov approximation to be made, resulting in local equations of motion for the averages of the chosen fields. While the general formalism is applicable to any set of variables, if the $\bm{A}(\bm{r})$ fields are not slowly varying one must solve a more complex set of equations for the $\bm{\phi}(\bm{r},t)$ fields.

The introduction of an auxiliary local density operator to enforce the constraint that the exact nonequilibrium and local density averages of the chosen fields are equal has important implications for the computation of the transport properties that enter the evolution equations. In particular, it allows one to bypass sampling from the exact nonequilibrium density, which may be difficult to carry out. The price paid for this simplification is the need to solve equations to determine the constraint fields. Nevertheless, the resulting equations are amenable to solution using various approximations and provide an alternate route to evaluating the microscopic expressions for transport properties under nonequilibrium conditions.

Although we have illustrated the formalism by studying a system of reactive quantum particles constrained to lie out of chemical equilibrium by coupling to chemostats, the general formalism can be applied to other systems. The choice of the degrees of freedom to be treated classically depends on the physical context, as does the manner in which the system is driven out of equilibrium. For instance, in molecular species, electronic as well as a selection of high-frequency nuclear modes could be treated quantum mechanically while others may be taken to be classical; if the solvent is a polar molecular fluid, collective polarization densities can be included in the description. The system may be driven out of equilibrium by external fields instead of chemostats and the transport can be studied by coupling to different thermostats, treated classically or quantum mechanically.

We have presented the formalism for systems whose evolution is governed by the quantum-classical Liouville equation. However, since this equation reduces to quantum evolution when the classical degrees of freedom are absent and to classical evolution if there is no quantum subsystem, the formalism also embodies these two limits.

\section*{Acknowledgements}
Financial support was received from the Natural Sciences and Engineering Research Council of Canada (JS,RK) and the Leverhulme Trust, RPG-2023-078 (RK).

\appendix

\section{Stress tensor and energy Flux}\label{app:stress-heat}

In the small gradient limit where the coarse-graining length is long compared to microscopic length scales, $\ell_\Delta \gg \ell_{\rm int}$, \cite{RSK24}
the full expressions for the classical and quantum components of the stress tensor $\hat{\bm{\tau}}=\bm{\tau}_c+\hat{\bm{\tau}}_q$ are
\begin{align*}
\bm{\tau}_c(\bm{r}) &= \sum_{i=1}^N\Big[ \frac{\bm{P}_i\bm{P}_i}{M} -\frac{1}{2} \sum_{j \neq i}^N 
\bm{R}_{ij} \partial_{\bm{R_{ij}}} U_C(R_{ij}) \Big]\Delta(\bm{R}_{i}-\bm{r})  \\
\hat{\bm{\tau}}_q(\bm{r}) &= -\frac{1}{2} \sum_{i=1}^N \sum_{ j \neq i}^N \bm{R}_{ij} \partial_{\bm{R_{ij}}} \hat{V}_{q}(R_{ij}) \Delta(\bm{R}_{i}-\bm{r}),
\end{align*}
and the energy flux, $\hat{\bm{\jmath}}_e= \bm{j}_{ec} + \hat{\bm{\jmath}}_{eq}$, contributions are
\begin{align*}
\bm{j_{ec}}(\bm{r}) &= \sum_{i=1}^N H_{ci} \frac{\bm{P}_i}{M} \Delta (\bm{R}_i-\bm{r})\nonumber \\
&- \frac{1}{2}\sum_{i=1}^N 
 \sum_{j \neq i} \bm{R}_{ij} \partial_{\bm{R_{ij}}} U_C(R_{ij}) \cdot \frac{\bm{P}_i}{M} \Delta (\bm{R}_i-\bm{r})\nonumber \\
\hat{\bm{\jmath}}_{eq}(\bm{r}) &= \sum_{i=1}^N \Theta_i^q \hat{H}_{qci} \frac{\bm{P}_i}{M} \Delta (\bm{R}_i-\bm{r}) \nonumber \\
&- \frac{1}{2} \sum_{i=1}^N \sum_{j \neq i}^N \bm{R}_{ij} \partial_{\bm{R_{ij}}} \hat{V}_{q} ({R}_{ij}) \cdot \frac{\bm{P}_i}{M} \Delta (\bm{R}_i-\bm{r}).
\end{align*}
where $\hat{V}_{q}({R}_{ij})$  and $U_c(R_{ij})$ are defined in Eqs.~(\ref{eq:int-pot}) and~(\ref{eq:classical-potential}).

Keeping only $\hat{V}_{qb}$ contributions to the interaction potential $\hat{V}_q$ for a dilute solution, their matrix elements in the adiabatic basis are diagonal for the classical fluxes $\bm{\tau}^{\bm{\ell}' \bm{\ell}}_c(\bm{r}) = \bm{\tau}_c(\bm{r}) \delta_{\bm{\ell}' \bm{\ell}}$, $\bm{j}_{ec}^{\bm{\ell}' \bm{\ell}}(\bm{r}) =
\bm{j}_{ec}(\bm{r}) \delta_{\bm{\ell}' \bm{\ell}}$, while the quantum fluxes have off-diagonal contributions,
\begin{eqnarray*}
{\bm{\tau}}^{\bm{\ell}' \bm{\ell}}_q(\bm{r}) &=& -\sum_{i=1}^N \sum_{j=1}^N \Theta_{i}^{b}\Theta_{j}^{q} \bm{R}_{ij} \Big( \partial_{\bm{R_{ij}}} {V}_{qb,\ell' \ell}(R_{ij})  \\
&& +\sum_{\ell`` } \big(  \bm{d}_{j,\ell' \ell''} (\bm{R}_{bj}){V}_{qb,\ell'' \ell}(R_{ij}) \nonumber \\
&&-{V}_{qb,\ell' \ell''}(R_{ij}) \bm{d}_{j,\ell'' \ell}(\bm{R}_{bj})  \big) \Big)\Delta(\bm{R}_{i}-\bm{r})  \delta^{(j)}_{\bm{\ell}' \bm{\ell}} ,  \nonumber
\end{eqnarray*}
and
\begin{align*}
{\bm{j}}_{eq}^{\bm{\ell}' \bm{\ell}}(\bm{r}) &= 
\sum_{i=1}^N \Theta_i^q \epsilon_{\ell i}(\bm{R}_{bi}) \frac{\bm{P}_i}{M} \Delta (\bm{R}_i-\bm{r}) \delta_{\bm{\ell}', \bm{\ell} 
}\nonumber \\ 
&-\sum_{i=1}^N \sum_{j=1}^N \Theta_{i}^{b}\Theta_{j}^{q} \bm{R}_{ij} \Big( \partial_{\bm{R_{ij}}} {V}_{qb,\ell' \ell}(R_{ij})  \\
& +\sum_{\ell'' } \big(  \bm{d}_{j,\ell' \ell''} (\bm{R}_{bj}){V}_{qb,\ell'' \ell}(R_{ij}) \nonumber \\
&-{V}_{qb,\ell' \ell''}(R_{ij}) \bm{d}_{j,\ell'' \ell}(\bm{R}_{bj})  \big) \Big) \cdot \frac{\bm{P}_i}{M} \Delta(\bm{R}_{i}-\bm{r})  \delta^{(j)}_{\bm{\ell}' \bm{\ell}}, \nonumber
\end{align*}
where $\delta^{(i)}_{\bm{\ell}'
\bm{\ell}}=\prod_{\substack{j=1\\j\ne i}}^{N} \Theta_j^q \delta_{{\ell}'j {\ell}j}$ and
$\bm{d}_{j,\ell' \ell''} (\bm{R}_{bj}) = \langle \ell' j; \bm{R}_{bj} | \partial_{\bm{R_{bj}}} | \ell'' j; \bm{R}_{bj}\rangle$ is the non-adiabatic matrix element that couples the $\ell'$ and $\ell''$ adiabatic states of quantum particle $j$.

In the adiabatic representation, the Kubo-like transform defined in Eq.~(\ref{eq:noneqKubo}) of the off-diagonal elements of the quantum fluxes $\bm{j}_q^{\bm{\ell}' \bm{\ell}}$, where $\bm{\ell} \neq \bm{\ell}'$ and $\hat{\bm{\jmath}}_q = \hat{\bm{\tau}}_{q}$ or $\hat{\bm{\jmath}}_q = \hat{\bm{\jmath}}_{eq}$, is related to the matrix elements by
\begin{equation*}
\overline{\bm{j}}_q^{\bm{\ell}' \bm{\ell}} = \left( \frac{1-e^{-\beta (\epsilon_{\bm{\ell}} - \epsilon_{\bm{\ell}'})}}{\beta (\epsilon_{\bm{\ell}} - \epsilon_{\bm{\ell}'})} \right) \bm{j}_q^{\bm{\ell}' \bm{\ell}}.
\end{equation*}
The Kubo-like-transformed fluxes only differ significantly from the untransformed fluxes if the gap between the adiabatic energies is small relative to the local thermal energy.  Nonetheless, since the nonadiabatic coupling matrix element
\begin{align*}
\bm{d}_{j,\ell \ell'}(\bm{R}_{bj})  &= \frac{- \langle \ell j; \bm{R}_{bj} | \partial_{\bm{R_{bj}}} \hat{V}_{Ij}| \ell'j; \bm{R}_{bj}\rangle}{\epsilon_\ell(\bm{R}_{bj}) - \epsilon_{\ell'} (\bm{R}_{bj})}
\end{align*}
is also large when the energy gap is small, the Kubo-like-transform cannot be neglected.

\section{Generalized equations of motion}\label{app:a-eqs-full}
The average values of the fluxes can be computed by replacing $\bm{P}_{i}$ by $\bm{P}_{i}^{+}=\bm{P}_{i}-M\bm{v}(\bm{R}_i)$ to obtain
\begin{eqnarray*}
\langle \hat{\mathcal{J}}_\gamma (\bm{r}) \rangle_t &=& -\partial_{\bm{r}}\cdot
\big( n_\gamma \bm{v} \big) -\nu_\gamma \langle j_R \rangle_t \nonumber \\
\langle {\mathcal{J}}_\rho (\bm{r}) \rangle_t &=&-\partial_{\bm{r}}\cdot
(\rho \bm{v}) \nonumber \\
\langle \hat{\bm{\mathcal{J}}}_g (\bm{r}) \rangle_t &=&  -\partial_{\bm{r}} \cdot \big(\rho\bm{v}\bm{v}) +\langle{}\bm{\tau}^+\rangle_{t}\big)\nonumber \\
\langle {\mathcal{J}}_{e^+} (\bm{r}) \rangle_t&=&-\partial_{\bm{r}} \cdot \big( \bm{v}  e^+ + \bm{v} \cdot \langle{}\bm{\tau}^{+}\big)
\rangle_{t} ) .
\end{eqnarray*}
Making use of the approximations discussed in the text and the expressions for the average fluxes, the evolution equations can now be written as
\begin{align*}
\partial_t n_\gamma &=
-\partial_{\bm{r}}\cdot \big( n_\gamma \bm{v} \big) -\nu_\gamma \langle j_R \rangle_t
\nonumber \\
& \quad - \nu_\gamma \beta {L}_{R} \ast  {\mathcal{A}}
+\nu_\gamma  \bm{L}_{R \gamma'} \ast \partial_{\bm{r'}} ( \beta \widetilde{\mu}_{\gamma'}) \nonumber \\
& \quad + \nu_\gamma  \bm{L}_{R v} \oast \partial_{\bm{r'}} (\beta\bm{v}) -\nu_\gamma  \bm{L}_{R e^+} \ast \partial_{\bm{r}'} \beta
\nonumber \\
& \quad - \partial_{\bm{r}} \cdot \bm{L}_{\gamma R}\ast {\mathcal{A}}
+ \partial_{\bm{r}} \cdot\bm{L}_{\gamma \gamma'} \ast \partial_{\bm{r'}} ( \beta \widetilde{\mu}_{\gamma'} ) \nonumber \\
& \quad + \partial_{\bm{r}} \cdot  \bm{L}_{\gamma v} \oast  \partial_{\bm{r'}} \big( \beta \bm{v} \big) -\partial_{\bm{r}} \cdot \bm{L}_{\gamma e^+} \ast \partial_{\bm{r}'} \beta \nonumber \\
\partial_t  \rho &=  -\partial_{\bm{r}}\cdot (\rho \bm{v})
\nonumber \\
\rho \partial_t \bm{v}  &=
-\rho \bm{v} \, \partial_{\bm{r}} \cdot \bm{v} - \partial_{\bm{r}} \cdot \bm{P}_h \nonumber \\
& \quad  - \partial_{\bm{r}}\cdot \bm{L}_{v R} \ast {\mathcal{A}} +\partial_{\bm{r}}\cdot \bm{L}_{v \gamma'} \ast \partial_{\bm{r'}} \big( \beta \widetilde{\mu}_{\gamma'} \big)
\nonumber \\
&   \quad +\partial_{\bm{r}}\cdot \bm{L}_{v v} \oast \beta \big( \partial_{\bm{r'}}\bm{v} \big) -\partial_{\bm{r}} \cdot
\bm{L}_{v e^+} \ast \partial_{\bm{r}^\prime} \beta \nonumber \\
\partial_t e^{+} &=
-\partial_{\bm{r}} \cdot \big(\bm{v} e^+ \big) - \bm{P}_h : \big( \partial_{\bm{r}} \bm{v} \big)\nonumber \\
& \quad
-\partial_{\bm{r}} \cdot
\bm{L}_{e^+ R} \ast {\mathcal{A}}  - \partial_{\bm{r}} \bm{v} : \bm{L}_{vR} \ast \mathcal{A} \nonumber   \\
&  \quad + \partial_{\bm{r}} \cdot \bm{L}_{e^+ \gamma'} \ast \partial_{\bm{r'}}
( \beta \widetilde{\mu}_{\gamma'} )
+ \partial_{\bm{r}} \cdot \bm{L}_{e^+ v} \oast  \beta \big( \partial_{\bm{r'}} \bm{v} \big)\nonumber \\
&\quad + \big( \partial_{\bm{r}} \bm{v} \big) : \bm{L}_{v \gamma} \ast \partial_{\bm{r'}} \big( \beta \widetilde{\mu}_\gamma \big) \nonumber \\
 & \quad + \big(\partial_{\bm{r}} \bm{v} \big) : \bm{L}_{vv} \oast \beta \big( \partial_{\bm{r'} }\bm{v} \big) - \partial_{\bm{r}} \cdot \bm{L}_{e^+ e^+} \ast \partial_{\bm{r}'} \beta \nonumber,
\end{align*}
where $e^+(\bm{r},t)  = e(\bm{r},t) - \frac{1}{2} \rho(\bm{r},t) v^{2}(\bm{r},t)$. In these equations $\rho(\bm{r},t) = M n(\bm{r},t)$ is the mass density, $\mathcal{A}(\bm{r},t)= \mu_B(\bm{r},t)-\mu_A(\bm{r},t)$ is the chemical affinity, and $\bm{P}_h (\bm{r},t) = \langle{}\bm{\tau{}}^{+}(\bm{r})\rangle_{t}$ is the hydrostatic pressure tensor.
In the equations above, $\bm{T} \oast \bm{U} \equiv \int d\bm{r}' \, \bm{T}_{\alpha \beta}(\bm{r}') \bm{U}_{\alpha \beta}(\bm{r}')$ denotes the contraction of second-rank tensors and an integration over the spatial argument $\bm{r}'$, and $\bm{T} : \bm{U} \equiv \bm{T}_{\alpha \beta}(\bm{r}) \bm{U}_{\alpha \beta}(\bm{r})$ is the contraction of the tensor indices alone.

\bibliography{qneqreact-refs}

\begin{thebibliography}{60}
\bibitem [{ex-()}]{ex-fields}%
  \BibitemOpen
  \href@noop {} {}\bibinfo {note} {If the system is subjected to time dependent
  external fields, these can be included in the quantum-classical Liouville
  operator. This generalization does not significantly change the structure of
  the formalism presented in this section.}\BibitemShut {Stop}%
\bibitem [{cla()}]{class-coord}%
  \BibitemOpen
  \href@noop {} {}\bibinfo {note} {More generally, if for a quantum molecule,
  in addition to its center of mass, some of its internal nuclear degrees of
  freedom may be treated classically, they can be included in the set of
  classical variables with little change in the structure of the equations
  derived in this paper.}\BibitemShut {Stop}%
\bibitem [{ort()}]{orthogonal}%
  \BibitemOpen
  \href@noop {} {}\bibinfo {note} {In some circumstances it may be convenient
  to orthogonalize the set of variables and operators in
  $\widehat{\bm{A}}(\bm{r})$ so that the matrix $\langle
  \widetilde{\bm{A}}(\bm{r}_1) \overline{\bm{A}}(\bm{r}_2)\rangle_t$ is
  diagonal. This simplifies the computations involving the inverse of this
  matrix.}\BibitemShut {Stop}%
\end{thebibliography}

\end{document}